\documentclass[journal]{IEEEtran}
\usepackage{graphicx,booktabs,fancyhdr,amsmath,amsfonts}
\usepackage{cite,bm,amssymb,amsthm,url,multirow,enumerate,verbatim}
\newcommand{\vb}{\boldsymbol}
\renewcommand{\tt}{\texttt}
\renewcommand{\it}{\textit}
\newtheorem{definition}{Definition}[section]

\begin{document}
\title{Requirements for Secure Clock Synchronization}  
\author{Lakshay~Narula,~\IEEEmembership{Student~Member,~IEEE,}
and~Todd~E.~Humphreys,~\IEEEmembership{Member,~IEEE}%
\thanks{L. Narula is with the Department of Electrical and Computer Engineering, Cockrell School of Engineering, The University of Texas at Austin, Austin, TX 78712 USA (email: lakshay.narula@utexas.edu).}%
\thanks{T. E. Humphreys is with the Department of Aerospace Engineering and Engineering Mechanics, Cockrell School of Engineering, The University of Texas at Austin, Austin, TX 78712 USA.}}
\maketitle
\begin{abstract} 
  This paper establishes a fundamental theory of secure clock synchronization.
  Accurate clock synchronization is the backbone of systems managing power
  distribution, financial transactions, telecommunication operations, database
  services, etc. Some clock synchronization (time transfer) systems, such as
  the Global Navigation Satellite Systems (GNSS), are based on one-way
  communication from a master to a slave clock. Others, such as the Network
  Transport Protocol (NTP), and the IEEE 1588 Precision Time Protocol (PTP),
  involve two-way communication between the master and slave.  This paper shows
  that all one-way time transfer protocols are vulnerable to replay attacks
  that can potentially compromise timing information. A set of conditions for
  secure two-way clock synchronization is proposed and proved to be necessary
  and sufficient.  It is shown that IEEE 1588 PTP, although a two-way
  synchronization protocol, is not compliant with these conditions, and is
  therefore insecure. Requirements for secure IEEE 1588 PTP are proposed, and a
  second example protocol is offered to illustrate the range of compliant
  systems.
\end{abstract}

\newif\ifpreprint
\preprintfalse

\ifpreprint

\pagestyle{plain}
\thispagestyle{fancy}  
\fancyhf{} 
\renewcommand{\headrulewidth}{0pt}
\rfoot{\footnotesize \bf May 2018 preprint of paper as accepted for publishing.} \lfoot{\footnotesize \bf
  Copyright \copyright~2018 by Lakshay Narula \\ and Todd E. Humphreys}

\else

\thispagestyle{empty}
\pagestyle{empty}

\fi


\section{Introduction}
\label{sec:introduction}
Secure clock synchronization is critical to a host of technologies and
infrastructure today. The phasor measurement units (PMUs) that enable
monitoring and control in power grids need timing information to synchronize
measurements across a wide geographical area \cite{phadke1994synchronized}.
Wireless communication networks synchronize their base stations to enable call
handoff \cite{mcneff2002global}.  Financial networks transfer time across the
globe to ensure a common time for pricing and transaction time-stamping
\cite{angel2014finance}. Cloud database services such as Google's Cloud
Spanner similarly require precise synchronization between the data centers to
maintain consistency~\cite{corbett2013spanner}. These clock synchronization
applications have sub-millisecond accuracy and stringent security
requirements.

Clock synchronization is performed either by over-the-wire packet-based
communication (NTP, PTP, etc.), or by over-the-air radio signals (GNSS
\cite{mcneff2002global}, cellular signals, LORAN \cite{shapiro1968time}, DCF77
\cite{bauch2009time}, etc.); both wired and wireless clock synchronization are
used extensively.  Synchronization by GNSS is the method of choice in systems
with the most stringent accuracy requirements. Equipped with atomic clocks
synchronized to the most accurate time standards available, GNSS satellites
can synchronize any number of stations on Earth to within a few tens of
nanoseconds \cite{allan1980accurate}. NTP is usually only accurate to a few
milliseconds, but essentially comes for free whenever the host device is
connected to a network.

One-way clock synchronization protocols are based on unidirectional
communication from the time master station, \tt{A}, to the slave station,
\tt{B}.  In such protocols, \tt{A} acts as a broadcast station and may send
out timing signals either continuously or periodically. The principal drawback
of one-way wireless clock synchronization protocols is their vulnerability to
delay attacks in which a man-in-the-middle (MITM) adversary nefariously
delays or repeats a valid transmission from one station to another.
Cryptographic and other measures can improve the security of one-way protocols
against delay and other signal- and data-level spoofing attacks
\cite{psiaki2016gnssSpoofing, wesson2018pincer, chou2014robust}, but, as will
be shown, such protocols remain fundamentally insecure because of their
inability to measure round trip time.  They can be secured against
unsophisticated attacks, but remain vulnerable to more powerful adversaries.

Two-way clock synchronization protocols involve bi-directional communication
between stations \tt{A} and \tt{B}. Such protocols enable measurement of the
round trip time of the timing signal, which is shown to be necessary for
detecting MITM delay attacks. This measurement, however, is not by itself
sufficient for provable security against such attacks.

This paper establishes a fundamental theory of secure clock synchronization.
In contrast to the current literature on timing security
\cite{ullmann2009delay, mizrahi2012game, moussa2016detection, yang2013time,
  tournier2009strategies, bhamidipati2016multi, ng2016robust}, the problem is
formalized with definitions, explicit assumptions, and proofs.  The major
contributions of this work are as follows:
\begin{enumerate}
\item One-way synchronization protocols are shown to be insecure against a
  MITM delay attack. Adversarial delay is shown to be indistinguishable
  from clock bias, and hence is unobservable without further assumptions.
\item A set of necessary conditions for secure two-way clock synchronization is
    presented and proved. Similar protocol-specific conditions have been
    previously proposed \cite{ullmann2009delay, moussa2016detection,
    annessi2017securetime}, but have not been generalized to apply to a
    universal clock synchronization model.
  \item The proposed necessary conditions, with stricter upper bounds, are
    shown to be sufficient for secure synchronization in presence of a
    probabilistic polynomial time (PPT) adversary. Provable security for clock
    synchronization has not previously been explored in the literature.
\item The two-way synchronization scheme of IEEE 1588 PTP is shown to violate a
    necessary condition for security. This is a known vulnerability of PTP for
    which a fix has been proposed \cite{ullmann2009delay}. Having established a
    theory for security, this paper is able to show that the proposed fix is
    sufficient but is not the minimal necessary modification. A more
    parsimonious security requirement for PTP is presented that is both
    necessary and sufficient for secure synchronization.
\item A generic construction of a secure two-way clock synchronization protocol
    is presented to illustrate the general applicability of the proposed
    necessary and sufficient conditions to a range of underlying protocols.
\end{enumerate}

This paper is a significant extension of
\cite{narula2016RequirementsSecurePlans}, by the same authors: (1) the
necessary conditions for security have been revamped to incorporate both
continuous and packet-based clock synchronization systems, (2) a sufficiency
proof for the security conditions has been formulated, and (3)
protocol-specific countermeasures presented in the literature have been unified
with the proposed conditions.

Wired clock synchronization is inherently more secure than its wireless
counterpart because physical access to cables is easier controlled than access
to radio channels. This paper primarily focuses on the more challenging task
of clock synchronization over a wireless channel; nonetheless, the attacks and
security protocols discussed herein also apply to wireline clock
synchronization protocols in the case where the adversary gets access to the
channel. For example, if an adversary is able to hijack a boundary clock in a
wireline PTP network, then the resulting vulnerabilities are equivalent to
that of wireless synchronization where the adversary has open access to the
radio channel. In fact, an adversarial boundary clock is even more potent than
a wireless adversary since it can completely block the authentic signal from
reaching \tt{B}.

The rest of this paper is organized as follows. Previous works on secure clock
synchronization, and their relation to this paper, are summarized in
Section~\ref{sec:related}. Section~\ref{sec:model} presents a generic model for
clock synchronization and shows that all possible one-way synchronization
protocols are insecure. Section~\ref{sec:necessary} presents the set of
security conditions for a wireless clock synchronization protocol, proving
these to be necessary by contradiction.  Section~\ref{sec:sufficient} presents
a proof of sufficiency for the same set of conditions with stricter upper
bounds. A construction of an example secure protocol is presented in
Section~\ref{sec:construction}, along with the security requirements for IEEE
1588 PTP.  Section~\ref{sec:simulation} presents a simulation study of a secure
clock synchronization model operating over a simplistic channel model.
Concluding remarks are made in Section~\ref{sec:conclusion}.

\section{Related Work}
\label{sec:related}
GNSS, NTP, and PTP are the most widely used protocols for clock
synchronization. A number of research efforts have been made to assess and
improve the security of these protocols. This section reviews some of the
notable efforts in the literature.

The GNSS jamming and spoofing threat has been recognized in the literature for
more than a decade. A survey of the current state-of-the-art in spoofing and
anti-spoofing techniques is presented in \cite{psiaki2016gnssSpoofing}. Recent
works on GNSS anti-spoofing techniques have specifically focused on the case of
timing security. Collaborative multi-receiver \cite{bhamidipati2016multi} and
direct time estimation \cite{ng2016robust} techniques have been proposed for
robust GNSS clock synchronization.

The growing popularity of IEEE 1588 PTP for synchronization in critical
infrastructure has brought about concerns regarding its security
\cite{ullmann2009delay, mizrahi2012game, moussa2016detection, yang2013time,
tournier2009strategies}.  The threats to IEEE 1588 PTP can broadly be
categorized into data-level attacks (such as modification of time messages) and
physical layer attacks (such as replay and delay attacks). While cryptographic
protocols are able to foil data-level attacks against realistic adversaries,
some signal-level attacks, such as the delay attack, remain open
vulnerabilities.  Unfortunately, their execution is relatively simple.
Signal-level attacks, such as the man-in-the-middle attack, have been studied in
the recent past. However, these studies
only include a brief discussion on countermeasure techniques, and no proof
or theoretical guarantee of the efficacy of the countermeasures has been
provided.

Ullman et al. \cite{ullmann2009delay} propose measuring the propagation delays
during initialization of clock synchronization and monitoring the propagation
delays during the normal operation of the time synchronization
protocol. However, \cite{ullmann2009delay} does not prove that such a defense
would be sufficient to prevent the delay attacks.

In \cite{moussa2016detection}, it is remarked that the clock offset computed
between multiple master clocks over a symmetric channel must be zero, and
thus, if multiple master clocks are available, they can detect any malicious
delay introduced by an adversary. However, this defense does not consider the
possibility that the adversary may only delay the packets sent to the slave
nodes.

The work presented in \cite{annessi2017securetime} is perhaps in closest
relation to the current paper. Annessi et al. upper bound the clock drift
between subsequent synchronization signals using a drift model, and perform
two-way exchange of timestamps such that the master clock is able to verify
the time at the slave.  Furthermore, given the maximum clock drift rate and
the maximum and minimum propagation delay of the timing signal, they derive an
upper bound on the adversarial delay that can go unnoticed. However, with
conservative bounds on the maximum clock drift rate and the variation in path
delays, the accuracy guarantees derived in \cite{annessi2017securetime} may be
insufficient for certain applications. Moreover, as will be shown in this
paper, they fail to take account of one the necessary conditions for secure
synchronization.

This paper abstracts the clock synchronization model and assesses its security
in a generic setting. It is shown that specialization of the generic security
conditions to the particular protocols assessed in the aforementioned efforts
leads to solutions similar or identical to those previously advanced. Thus,
establishing the fundamental theory of secure clock synchronization also
serves to unify the prior work in the literature.

\section{System Model}
\label{sec:model}
\begin{figure}[ht]
\centering
\includegraphics[width = 8.5cm]{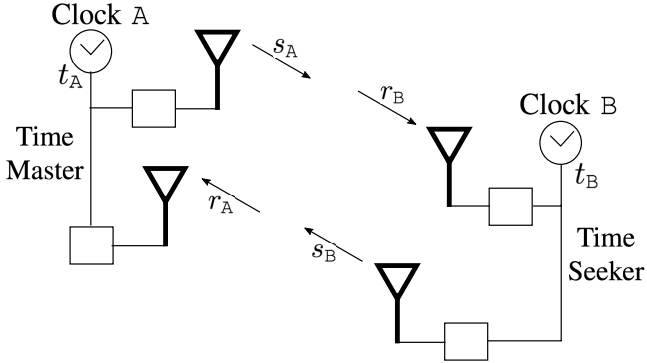}
\caption{Abstract model of a clock synchronization system with a time master
station \tt{A} and a time seeker station \tt{B}. The antenna outputs are driven
by the clock through the receiver and transmitter blocks.}
\label{fig:sysmodel}
\end{figure}

\begin{table}[tb]\caption{Notation used in this paper}
    \centering
    \begin{tabular}{c|p{7cm}}
        \toprule
        \tt{A}                                      &   Time master station \\
        \midrule
        \tt{B}                                      &   Time seeker station \\
        \midrule
        $t_\tt{m}^{\tt{m}_i}$                       &   Transmit time, according to \tt{m}, of its $i$th signal feature \\
        \midrule
        $t_\tt{n}^{\tt{m}_i}$                       &   Receipt time, according to \tt{n}, of the $i$th signal feature transmitted by \tt{m}    \\
        \midrule
        $\tau_\tt{mn}^i$                            &   Delay, in true time, experienced by the $i$th feature in propagating from \tt{m} to \tt{n} \\
        \midrule
        $\tau_{\tt{mn}_\mathcal{M}}^i$              &   Component of $\tau_\tt{mn}^i$ introduced by the man-in-the-middle adversary \\
        \midrule
        $\tau_{\tt{mn}_\mathcal{N}}^i$              &   Component of $\tau_\tt{mn}^i$ due to natural factors, including processing, transmission, and propagation delay \\
        \midrule
        $\bar{\tau}_\tt{mn}^i$                      &   Modeled or \emph{a priori} estimate of $\tau_{\tt{mn}_\mathcal{N}}^i$   \\
        \midrule
        $\tilde{\tau}_{\tt{mn}_\mathcal{N}}^i$      &   $\tau_{\tt{mn}_\mathcal{N}}^i - \bar{\tau}_\tt{mn}^i$   \\
        \midrule
        $\tau_\texttt{BB}$                          &   Delay, in true time, between the receipt of \emph{sync} and transmission of \emph{response} at \tt{B}   \\
        \midrule
        $\bar{\tau}_\texttt{BB}$                    &   Delay, according to \tt{B}, between the receipt of \emph{sync} and transmission of \emph{response} at \tt{B}    \\
        \midrule
        $\tilde{\tau}_\texttt{BB}$                  &   $\tau_\texttt{BB} - \bar{\tau}_\texttt{BB}$ \\
        \midrule
        $\Delta t_\tt{AB}^i$                        &   Clock offset between \tt{A} and \tt{B} at the time of receipt of the $i$th feature at \tt{B}    \\
        \midrule
        $\Delta \hat{t}_\tt{AB}^i$                  &   \tt{B}'s best estimate of $\Delta t_\tt{AB}^i$  \\
        \midrule
        $w_\tt{mn}^i$                               &   Measurement noise associated with the measured time-of-arrival of the $i$th signal feature from \tt{m} at \tt{n} \\
        \midrule
        $\tau_{\rm RTT}^{ij}$                       &   Round trip time, in true time, involving the $i$th and $j$th signal features of \tt{A} and \tt{B}, respectively \\
        \midrule
        $\bar{\tau}_{\rm RTT}^{ij}$                 &   Modeled or \emph{a priori} estimate of $\tau_{\rm RTT}^{ij}$    \\
        \midrule
        $z_{\rm RTT}^{ij}$                          &   A noisy measurement of $\tau_{\rm RTT}^{ij}$    \\
        \bottomrule
    \end{tabular}
    \label{tab:notation}
\end{table}
A general system model for clock synchronization is shown in Fig.
\ref{fig:sysmodel}.  The time seeker station, \tt{B}, wishes to synchronize
its clock to that of the time master station, \tt{A}. For wireless
synchronization applications, stations \tt{A} and \tt{B} are assumed to have
known locations, $\bm{x}_\tt{A}$ and $\bm{x}_\tt{B}$, respectively.  Due to
clock imperfections, the time at station \tt{B}, $t_\tt{B}$, continuously
drifts with respect to $t_\tt{A}$, the time at station \tt{A}. Station \tt{B}
seeks to track the relative drift of its clock by an exchange of signals
between \tt{A} and \tt{B}.  Without loss of generality, this paper assumes
$t_\tt{A}$ is equivalent to true time (relative to some reference epoch), a
close proxy for which is GPS system time.

It is assumed that \tt{A} and \tt{B} are able to exchange cryptographic keys
securely, if required. This exchange may occur over a public channel via a
protocol such as the Diffie-Hellman key exchange \cite{merkle1978secure} or via
quantum key exchange techniques \cite{bennett1984quantum, ekert1991quantum}.
Alternatively, symmetric keys for neighboring stations may be loaded at the
time of installation.

Station \tt{A} sends out a \it{sync} signal, $s_\tt{A}$, having distinct
features which can be disambiguated from one another by observing a window
of the signal containing the feature. The transition in $s_\tt{A}$ marking the
beginning of a data packet is an example of such a signal feature.
Furthermore, the system at \tt{A} is designed such that the $k$th feature is
transmitted at time $t_\tt{A}^{\tt{A}_k}$.  \tt{B} either knows
$t_\tt{A}^{\tt{A}_k}$ by prior arrangement, or a digital representation of
$t_\tt{A}^{\tt{A}_k}$ is encoded in $s_\tt{A}$ (e.g., a timestamp).  In any
case, \tt{B} knows when the $k$th feature was sent, according to \tt{A}'s
clock.  This sets up a bijection
\begin{equation}
    S_{\tt{A}}^k \rightleftharpoons k \rightleftharpoons t_\tt{A}^{\tt{A}_k}
    \label{eq:S_Ak_bijection_t_Ak}
  \end{equation}
where $S_{\tt{A}}^k$ represents a window of $s_\tt{A}$ containing the $k$th
feature.

Station \tt{B}'s received \it{sync} signal, denoted $r_\tt{B}$, is a delayed
and noisy replica of $s_\tt{A}$. Let $\tau_\tt{AB}^k$ denote the delay (in true
time) experienced by the $k$th feature of $s_\tt{A}$ as it travels from \tt{A}
to \tt{B}. For line-of-sight (LOS) wireless communication, $\tau_\tt{AB}^k$ is
the sum of the free-space propagation delay over the distance
$\|\bm{x}_\tt{B}-\bm{x}_\tt{A}\|$ and additional delays due to interaction of
the timing signal with the intervening channel.

\subsection{One-Way Clock Synchronization Model}
In one-way clock synchronization, the exchange of signals between \tt{A} and
\tt{B} terminates with reception of the \it{sync} signal at \tt{B}. Let
$t_\tt{B}^{\tt{A}_k}$ denote the time according to \tt{B} at which the $k$th
feature of $s_\tt{A}$ is received at \tt{B}. The window captured by \tt{B}
containing the $k$th feature of $s_\tt{A}$, denoted $R_\tt{B}^k$, enables
\tt{B} to measure $t_\tt{B}^{\tt{A}_k}$ to within a small error caused by
measurement noise. This error, $w_{\tt{AB}}^k$, is modeled as zero-mean with
variance $\sigma_\epsilon^2$. The measurement itself, denoted $z_{\tt{B}}^k$,
is modeled as
\begin{align}
    z_{\tt{B}}^k &= t_\tt{B}^{\tt{A}_k} + w_{\tt{AB}}^k \nonumber \\
    & = t_\tt{A}^{\tt{A}_k} + \tau_{\tt{AB}}^k - \Delta t_\tt{AB}^k + w_{\tt{AB}}^k 
\end{align}
where
\begin{equation}
    \Delta t_\tt{AB}^k \equiv t_\tt{A}^{\tt{A}_k} + \tau_{\tt{AB}}^k - t_\tt{B}^{\tt{A}_k}
    \label{eq:offset}
\end{equation}
is the unknown time offset \tt{B} wishes to estimate. As the bijection in
(\ref{eq:S_Ak_bijection_t_Ak}) is known to \tt{B}, \tt{B} can obtain
$t_\tt{A}^{\tt{A}_k}$ for the $k$th detected feature. If a prior estimate
$\bar{\tau}_{\tt{AB}}^k$ of the delay $\tau_{\tt{AB}}^k$ is available to
\tt{B}, then the desired time offset can be estimated as
\begin{align}
    \Delta \hat{t}_\tt{AB}^k &= t_\tt{A}^{\tt{A}_k} + \bar{\tau}_{\tt{AB}}^k - z_{\tt{B}}^k
    \label{eq:time_recovery}
\end{align}

As a concrete example, consider the case of clock synchronization via GNSS in
which $\tt{B}$ is a GNSS receiver in a known fixed location $\vb{x}_\tt{B}$,
and \tt{A} is a GNSS satellite whose location is known to vary with time as
$\vb{x}_\tt{A}(t_{\tt{A}})$. On detection of the $k$th feature in a window of
captured data, \tt{B} determines $t_\tt{A}^{\tt{A}_k}$ using
(\ref{eq:S_Ak_bijection_t_Ak}) and also makes the measurement
\begin{align*}
    z_{\tt{B}}^k &= t_\tt{A}^{\tt{A}_k} + \tau_{\tt{AB}}^k - \Delta t_\tt{AB}^k + w_{\tt{AB}}^k \\
    &= t_\tt{A}^{\tt{A}_k} + \left[ \frac{\|\bm{x}_\tt{B}-\bm{x}_\tt{A}(t_\tt{A}^{\tt{A}_k})\| + D_\rho^k}{c} \right] - \Delta t_\tt{AB}^k + w_{\tt{AB}}^k
\end{align*}
where $D_\rho^k$ is the sum of excess ionospheric and neutral-atmospheric
delays (in distance units) and $c$ is the speed of light.

The known receiver and satellite positions can be invoked to model the
signal's propagation delay as
\begin{align*}
    \bar{\tau}_{\tt{AB}}^k &= \frac{\|\bm{x}_\tt{B}-\bm{x}_\tt{A}(t_\tt{A}^{\tt{A}_k})\| + \bar{D}_\rho^k}{c}
\end{align*}
where $\bar{D}_\rho^k$ is a model of the excess delay $D_\rho^k$ at the time
of receipt of the $k$th feature at \tt{B}. The modeled excess delay is based
on atmospheric models possibly refined by dual-frequency measurements
\cite{p_misra06_smp}.  An estimate of the time offset,
$\Delta \hat{t}_\tt{AB}^k$, can then be made using $t_\tt{A}^{\tt{A}_k}$,
$z_{\tt{B}}^k$, and $\bar{\tau}_{\tt{AB}}^k$ in (\ref{eq:time_recovery}).

It must be noted that, for one-way clock synchronization, any errors in the
estimate of the distance between \tt{A} and \tt{B}, and in the estimate of the
excess channel delay, will appear as an error in the estimate of the time
offset.

\subsection{Two-Way Clock Synchronization Model}
As discussed above, if an estimate of $\bar{\tau}_{\tt{AB}}^k$ is available,
then clock synchronization is complete after \tt{B} receives the \it{sync}
signal $r_\tt{B}$. The \it{response} signal from \tt{B} in a two-way protocol
is typically used to either determine, or refine, the estimate of
$\bar{\tau}_{\tt{AB}}^k$ with a measurement of the round trip time (RTT). The
ability to measure RTT obviates the requirement that
$\|\bm{x}_\tt{B}-\bm{x}_\tt{A}\|$ be known \it{a priori}. In IEEE 1588 PTP,
for example, RTT is measured to initially obtain, and periodically refine, the
value of $\bar{\tau}_{\tt{AB}}^k$ used in deriving $\Delta \hat{t}_\tt{AB}^k$
from (\ref{eq:time_recovery}).

In the system model considered in this paper, station \tt{B} transmits a
\it{response} $s_\tt{B}$ that is designed such that (1) there is a one-to-one
mapping $l(k)$ between the $l$th feature in $s_\tt{B}$ and the $k$th feature
in $s_\tt{A}$, and (2) the $l$th feature's index can be inferred by
observation of a window containing it. Symbolically, if $S_\tt{B}^{l}$ is a
window of $s_\tt{B}$ containing the $l$th feature of the \it{response} signal,
then
\begin{align}
    S_\tt{B}^{l} \rightleftharpoons l(k) \rightleftharpoons k
    \label{eq:SBl_l_k_bijection}
\end{align}
On receipt of the $k$th feature in $s_\tt{A}$, at time $t_{\tt{B}}^{\tt{A}_k}$
by $\tt{B}$'s clock, but at $z_{\tt{B}}^k$ as measured by \tt{B},
\tt{B} transmits the $l$th feature in $s_\tt{B}$ after a short delay,
$\tau_\tt{BB}$ (in true time), hereon referred to as the \it{layover time}.

The layover time is introduced as a practical consideration. On receipt of
\tt{A}'s $k$th feature, \tt{B} is physically unable to transmit its own $l$th
feature with zero delay. Thus, \tt{B} is allowed to specify a short layover
time, $\bar{\tau}_\tt{BB}$, after which it intends to launch its $l$th feature.
It is important to note that the actual layover time, $\tau_\tt{BB}$, will not
be the same as the intended layover time due to (1) non-zero measurement noise
$w_\tt{AB}^k$ and (2) non-zero frequency offset of the clock at \tt{B} with
respect to true time. However, if the layover time is sufficiently short and
the measurement noise is benign, the difference $\bar{\tau}_\tt{BB} -
\tau_\tt{BB}$ can be made negligible compared to the time synchronization
requirement, with the actual value depending on the quality of \tt{B}'s clock.

Station \tt{A} receives the \it{response} signal as a delayed and noisy
replica of $s_\tt{B}$, denoted $r_\tt{A}$. The delay experienced by the $l$th
feature as it travels from \tt{B} to \tt{A}, in true time, is denoted
$\tau_\tt{BA}^l$. Station \tt{A} captures a window $R_\tt{A}^l$ of $r_\tt{A}$
that enables \tt{A} to identify the $l$th feature in $s_\tt{B}$ according to
(\ref{eq:SBl_l_k_bijection}), and to infer that the received feature is in
response to the $k$th feature transmitted by \tt{A}. Furthermore, \tt{A} makes
a noise-corrupted measurement $z_\tt{A}^l$ of the time-of-arrival of the $l$th
feature in $s_{\tt{B}}$, according to \tt{A}'s clock. The noise, denoted
$w_\tt{BA}^l$, is again modeled as zero-mean with variance
$\sigma_\epsilon^2$. The full measurement model is given by
\begin{align*}
    z_\tt{A}^l &= t_\tt{A}^{\tt{B}_l} + w_\tt{BA}^l \\
    &= t_\tt{A}^{\tt{A}_k} + \tau_\tt{AB}^k + \tau_\tt{BB} + \tau_\tt{BA}^l + w_\tt{BA}^l
\end{align*}
Since $t_\tt{A}^{\tt{A}_k}$ is exactly known at \tt{A}, a direct noisy
measurement of the round trip time $\tau_{\tt{AB}}^k + \tau_{\tt{BB}} +
\tau_{\tt{BA}}^l$ can be made as
\begin{equation}
    z_{\rm RTT}^{kl} \equiv z_{\tt{A}}^l - t_\tt{A}^{\tt{A}_k}
    \label{eq:RTT}
\end{equation}
Note that the noise $w_\tt{BA}^l$ and $w_\tt{AB}^k$ in $z_{\rm RTT}^{kl}$ is
embedded within $z_{\tt{A}}^l$ and $\tau_\tt{BB}$, respectively.  Under the assumption of
symmetric delays, i.e., $\tau_{\tt{AB}}^k = \tau_{\tt{BA}}^l$, and with
knowledge of $\bar{\tau}_{\tt{BB}}$, the measured RTT in (\ref{eq:RTT}) can be
exploited to improve the modeled propagation delay for future exchanges between
\tt{A} and \tt{B}:
\begin{equation*}
    \bar{\tau}_\tt{AB}^m = \bar{\tau}_\tt{BA}^n = \frac{z_{\rm RTT}^{kl} - \bar{\tau}_{\tt{BB}}}{2}
\end{equation*}
where $m > k$ and $n > l$.

\begin{figure}[ht!]
    \centering
    \includegraphics[width=9cm]{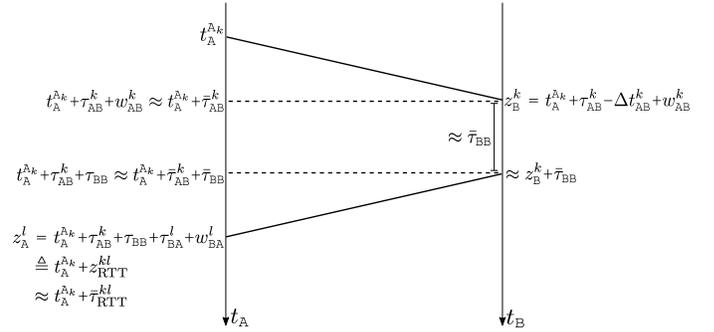}
    \caption{Two-way exchange of \emph{sync} and \emph{response} messages between \tt{A} and \tt{B} in the absence of a man-in-the-middle adversary.}
    \label{fig:time_transfer}
\end{figure}

The two-way exchange of \emph{sync} and \emph{response} messages is summarized
visually in Fig.~\ref{fig:time_transfer}.

Since RTT will play a central role in the discussion on secure
synchronization later on, various definitions and assumptions concerning RTT
are stated here for clarity:
\begin{itemize}
\item RTT for the $k$th feature in $s_\tt{A}$
  and the corresponding $l$th feature in $s_\tt{B}$ is defined as
  \[\tau_{\rm RTT}^{kl} \equiv \tau_\tt{AB}^k+\tau_\tt{BB}+\tau_\tt{BA}^l \]
\item Measured RTT includes, in addition to RTT, measurement noise at
    \tt{A}; it is modeled as
    \[
        z_{\rm RTT}^{kl} = \tau_\tt{AB}^k+\tau_\tt{BB}+\tau_\tt{BA}^l+w_\tt{BA}^l
    \]
\item Modeled RTT, also called the prior estimate of RTT, is defined as
    \begin{equation}
        \bar{\tau}_{\rm RTT}^{kl} \equiv \bar{\tau}_\tt{AB}^k+\bar{\tau}_\tt{BB}+\bar{\tau}_\tt{BA}^l
        \label{eq:tRTT_bar}
    \end{equation}
    For example, in the case of wireless clock synchronization with LOS electromagnetic
    signals, a prior estimate of RTT is based on the distance between \tt{A}
    and \tt{B} and on models of channel delays in excess of free-space
    propagation between these.
\item The modeled RTT, $\bar{\tau}_{\rm RTT}^{kl}$, can be refined with
    measurements of RTT in a two-way protocol. Alternatively, as will be
    discussed later, if an accurate modeled RTT is available, it and the
    measured RTT can be used to detect delay attacks.
\item Unambiguous measurement of RTT requires that there exist a one-to-one
    mapping between the signal features in $s_\tt{A}$ and $s_\tt{B}$, as
    mathematically represented in (\ref{eq:SBl_l_k_bijection}). On detection
    of the $l$th feature in $s_\tt{B}$, \tt{A} must be able to deduce that
    this feature was transmitted approximately $\bar{\tau}_\tt{BB}$ after
    \tt{B} received the $k$th feature in $s_\tt{A}$. This requirement is
    appropriately a part of the RTT definition since it enables \tt{A} to
    unambiguously measure RTT.
\end{itemize}

\subsection{Attack Model}
The attack model in this paper considers a MITM adversary $\mathcal{M}$. The
available computational resources allow $\mathcal{M}$ to execute probabilistic
polynomial time (PPT) algorithms. $\mathcal{M}$ can receive, detect, and
replay signals from \tt{A} and \tt{B} with arbitrarily precise directional
antennas.  Additionally, $\mathcal{M}$ has precise knowledge of
$\bm{x}_\tt{A}$ and $\bm{x}_\tt{B}$, and can take up any position around or
between the two stations. It has unrestricted access to the signals that
\tt{A} and \tt{B} exchange over the air, and has complete knowledge of their
synchronization protocol save for the cryptographic keys.

Let $L$ denote the alert limit, defined as the error in time synchronization
not to be exceeded without issuing an alert.
\begin{definition}
  Clock synchronization is defined to be compromised if
  ${\normalfont |\Delta t_\tt{AB} - \Delta \hat{t}_{\tt{AB}}| \geq L}$.
    \label{def:compromised}
\end{definition}

Note that, in the absence of an adversary, clock synchronization is not
compromised so long as
\begin{equation*}
    |\tau_\tt{AB}^k - \bar{\tau}_\tt{AB}^k + w_\tt{AB}^k| < L
\end{equation*}
However, in the presence of a MITM adversary, the \it{sync} signal is
delayed or advanced such that
\begin{equation}
    \tau_\tt{AB}^k = \tau_{\tt{AB}_\mathcal{N}}^k + \tau_{\tt{AB}_\mathcal{M}}^k
\end{equation}
where $\tau_{\tt{AB}_\mathcal{N}}^k > 0$ is the natural or physical delay
(equal to $\tau_\tt{AB}^k$ in the absence of an adversary) and
$\tau_{\tt{AB}_\mathcal{M}}^k \geq 0$ is the adversarial delay. In this case,
if
\begin{align}
    |\tau_\tt{AB}^k - \bar{\tau}_\tt{AB}^k + w_\tt{AB}^k| = 
    |\tau_{\tt{AB}_\mathcal{N}}^k - \bar{\tau}_\tt{AB}^k +
  \tau_{\tt{AB}_\mathcal{M}}^k + w_\tt{AB}^k| \geq L \label{eq:compromised}
\end{align}
then clock synchronization is compromised.

\subsection{Vulnerability of One-Way Clock Synchronization}
One-way clock synchronization is fundamentally vulnerable to a delay
attack because it provides no mechanism to measure RTT.  The adversary
$\mathcal{M}$ can compromise any one-way wireless clock synchronization
protocol by retransmitting the authentic \it{sync} signal from \tt{A} such
that the retransmitted signal, $s_\mathcal{M}$, overpowers or otherwise
supersedes the authentic signal $s_\tt{A}$.  In the absence of additional
assumptions beyond those underpinning the one-way protocol described earlier,
$\mathcal{M}$ can introduce an arbitrary delay $\tau_{\tt{AB}_\mathcal{M}}^k$
in its retransmission, thereby compromising the synchronization process.

Note that whereas counterfeit signal attacks can be prevented by
authentication and cryptographic methods \cite{wesson2011proposedNMA}, these
techniques do not prevent delay attacks because the delayed or repeated signal
has the same cryptographic characteristics as that of the genuine signal, the
only difference being that it is received with a (possibly small) additional
delay.

The delay introduced by $\mathcal{M}$ is added to the natural delay,
$\tau_{\tt{AB}_\mathcal{N}}^k$, of the signal between \tt{A} and \tt{B}. As a
result, an error of $\approx \tau_{\tt{AB}_\mathcal{M}}^k$ is introduced in
the estimated time offset at \tt{B}. From (\ref{eq:time_recovery}), it follows
that
\begin{align}
    \Delta \hat{t}_\tt{AB}^k &= t_\tt{A}^{\tt{A}_k} + \bar{\tau}_{\tt{AB}}^k - z_{\tt{B}}^k \notag \\
    &= t_\tt{A}^{\tt{A}_k} + \bar{\tau}_{\tt{AB}}^k - (t_\tt{A}^{\tt{A}_k} + \tau_{\tt{AB}}^k - \Delta t_\tt{AB}^k + w_\tt{AB}^k) \notag \\
    &= (\bar{\tau}_{\tt{AB}}^k - \tau_{\tt{AB}_\mathcal{N}}^k) - \tau_{\tt{AB}_\mathcal{M}}^k + \Delta t_\tt{AB}^k - w_\tt{AB}^k \notag \\
    &\approx \Delta t_\tt{AB}^k - \tau_{\tt{AB}_\mathcal{M}}^k \label{eq:error_adversary}
\end{align}
where it is assumed that the error due to inaccurately modeled delay is
negligible and that $\sigma_\epsilon \ll \tau_{\tt{AB}_\mathcal{M}}^k$. In the
absence of an RTT measurement, and without further assumptions on the nature
of the protocol or the clock drift model considered, the adversarial delay
$\tau_{\tt{AB}_\mathcal{M}}$ is indistinguishable from a clock offset of the
same magnitude.

To be sure, measures can be taken to make a MITM delay attack harder to
execute without detection. But, importantly, these measures cannot
guarantee that the synchronization will remain uncompromised.  Various measures
proposed in the literature, and their shortcomings, are discussed below.
\paragraph{Received Signal Strength Monitoring} The adversary $\mathcal{M}$
might attempt to overpower the authentic signal in order to spoof the
\it{sync} message, leading to an increase in the total signal power received
at \tt{B}.  Station \tt{B} could monitor the received signal strength (RSS) to
detect such an attack \cite{akos2012s}.  However, a potent adversary
could transmit, in addition to its delayed signal, an amplitude-matched,
phase-inverted nulling signal that annihilates the authentic \it{sync} signal
$s_{\tt{A}}$ as received at \tt{B}, thus preventing an unusual increase in
received power at \tt{B}.  If $\mathcal{M}$ is positioned along the
straight-line path between \tt{A} and \tt{B}, nulling of $s_{\tt{A}}$ can be
effected without prior knowledge of $s_{\tt{A}}$.  A laboratory
demonstration of such nulling is reported in \cite{humphreysGNSShandbook}.
\paragraph{Selective Rejection of False Signal} If \tt{B} receives both the
authentic and false (delayed) \it{sync} signals, it may be able to apply
angle-of-arrival or signal processing techniques to selectively reject the
delayed signal
\cite{psiaki2016gnssSpoofing,wesson2018pincer,meurer2012robust,borio2013panova}.
However, discrimination based on angle-of-arrival fails if $\mathcal{M}$ is
positioned along the line from \tt{A} to \tt{B}, and, as conceded in
\cite{wesson2018pincer}, signal-processing-based techniques for selective
rejection of false signals can be thwarted by an adversary transmitting an
additional nulling signal, as described above.
\paragraph{Collaborative Verification} Multiple time seekers may attempt to
synchronize to the same time master. In this scenario, the time seekers can
potentially detect malicious activity by cross-checking the received signals
\cite{bhamidipati2016multi}. In the simplest implementation, all time seekers
can collaborate to verify that they are synchronized amongst each other. In
case of an uncoordinated attack against a subset of time seekers, this
verification would expose the attack since the time offset computed at the
attacked subset would be different from that computed at the other
stations. In principle, however, it is possible for an adversary to execute a
coordinated attack against all the time seekers, thus concealing its presence.

\section{Necessary Conditions for Secure Synchronization}
\label{sec:necessary}
This section presents a set of conditions for secure two-way clock
synchronization and proves these to be necessary by contradiction.  In other
words, it is shown that if a two-way clock synchronization protocol does not
satisfy any one of these proposed conditions, there exists an attack that can
compromise clock synchronization without detection.

It is important to note that the ability to measure RTT in a two-way protocol
is necessary, but not sufficient, for provably secure synchronization.  As an
example, IEEE 1588 PTP is a two-way protocol that has been proposed as an
alternative to GNSS for sub-microsecond clock synchronization in critical
infrastructure such as the PMU network. But, despite the bi-directional
exchange between stations, and hence the ability to measure RTT, recent work
has shown that PTP is vulnerable to delay attacks in which a MITM introduces
asymmetric delay between \tt{A} and \tt{B}. Asymmetric delay breaks the
assumption that $\tau_\tt{AB}^k = \tau_\tt{BA}^k$ and leads to an erroneous
prior for $\bar{\tau}_\tt{AB}$ and $\bar{\tau}_\tt{BA}$ for future exchanges.
This vulnerability is documented in both the literature
\cite{ullmann2009delay, moussa2016detection, annessi2017securetime} and the
IEEE 1588-2008 standard.  Thus, a secure two-way clock synchronization
protocol must satisfy additional security requirements beyond the ability to
measure RTT.

The conditions introduced below are not tied to any specific protocol, unlike
some measures proposed in the current literature \cite{ullmann2009delay,
  mizrahi2012game, moussa2016detection, yang2013time, tournier2009strategies,
  bhamidipati2016multi, ng2016robust}. They are generally applicable to any
two-way protocol (e.g., PTP) for which the foregoing two-way synchronization
model applies.

Assuming the time master \tt{A} initiates the two-way communication, the
necessary conditions for secure clock synchronization are as follows:
\begin{enumerate}
\item Both \tt{A} and \tt{B} must transmit unpredictable waveforms to prevent
    the adversary $\mathcal{M}$ from generating counterfeit signals that pass
    authentication. In practice, this implies the use of a cryptographic
    construct such as a message authentication code (MAC) or a digital
    signature.
  \item The propagation time of the signal must be irreducible to within the
    alert limit $L$ along both signal paths. For wireless clock
    synchronization, this condition implies synchronization via LOS electromagnetic
    signals as $L \rightarrow 0$.
  \item The RTT between \tt{A} and \tt{B} must be known to \tt{A} and
    measurable by \tt{A} to within the alert limit $L$. The RTT must include
    the delays internal to both \tt{A} and \tt{B}, in addition to the
    propagation delay.  Station \tt{A} must know of any intentional delay
    introduced by \tt{B}, such as the layover time $\tau_\tt{BB}$ introduced
    earlier.
\end{enumerate}

\subsection{Proof of Necessity of Conditions}
\subsubsection{Stations {\normalfont \tt{A}} and {\normalfont \tt{B}} must
  transmit unpredictable signals}
To prove this condition is necessary, two scenarios are considered: \it{a)}
station \tt{A} transmits a signal waveform $s_\tt{A}$ that is predictable,
and, \it{b)} station \tt{B} transmits a signal waveform $s_\tt{B}$ that is
predictable.

\paragraph{{\normalfont $s_\tt{A}$} is predictable}
$\mathcal{M}$ can compromise synchronization without detection as follows:
\begin{enumerate}[i)]
\item $\mathcal{M}$ takes up a position between \tt{A} and \tt{B} along the
  line joining the antennas at the two stations.
\item $\mathcal{M}$ initially transmits a replica of $s_\tt{A}$ such that
  \tt{B} receives identical signals from both \tt{A} and $\mathcal{M}$.
  Subsequently, $\mathcal{M}$ increases its signal power or otherwise
  supersedes $s_\tt{A}$ (e.g., via signal nulling, as discussed earlier) such
  that \tt{B} tracks $s_\mathcal{M}$, the signal transmitted by $\mathcal{M}$.
  (Hereafter, whenever signals from $\mathcal{M}$ compete with those from
  \tt{A} or \tt{B}, it will be assumed that those from $\mathcal{M}$ exert
  control.)
\item Exploiting the predictability of $s_\tt{A}$, $\mathcal{M}$ advances its
  replica $s_\mathcal{M}$ with respect to $s_\tt{A}$ by
  $|\tau_{\tt{AB}_\mathcal{M}}^k|$, where $\tau_{\tt{AB}_\mathcal{M}}^k < 0$.
  \tt{B} tracks the advanced signal, resulting in an error of
  $\tau_{\tt{AB}_\mathcal{M}}^k$ in the computed $\Delta \hat{t}_\tt{AB}^k$ as
  shown in (\ref{eq:error_adversary}).
\item \tt{B} transmits the unpredictable \it{response} $s_\tt{B}$ compliant
  with the prearranged layover time $\bar{\tau}_\tt{BB}$. $\mathcal{M}$
  intercepts this signal from \tt{B}, and replays it to \tt{A} with a delay of
  $\tau_{\tt{BA}_\mathcal{M}}^l = -\tau_{\tt{AB}_\mathcal{M}}^k > 0$, causing
  \tt{A} to track the delayed signal.  As a result, the RTT is
  $\tau_\tt{AB}^k+\tau_\tt{BB}+\tau_\tt{BA}^l$ as \tt{A} expects. In summary:
    \begin{align*}
        \tau_\tt{AB}^k &= \tau_{\tt{AB}_\mathcal{N}}^k + \tau_{\tt{AB}_\mathcal{M}}^k \\
        \tau_\tt{BA}^l &= \tau_{\tt{BA}_\mathcal{N}}^l + \tau_{\tt{BA}_\mathcal{M}}^l = \tau_{\tt{BA}_\mathcal{N}}^l - \tau_{\tt{AB}_\mathcal{M}}^k \\
        \Rightarrow \tau_\tt{AB}^k + \tau_\tt{BA}^l &= \tau_{\tt{AB}_\mathcal{N}}^k + \tau_{\tt{BA}_\mathcal{N}}^l
    \end{align*}
    Thus, $\mathcal{M}$ undoes the effect of its \it{sync} advance, preventing
    \tt{A} from detecting the attack.
\end{enumerate}

\paragraph{{\normalfont $s_\tt{B}$} is predictable}
$\mathcal{M}$ can compromise synchronization without detection by replicating
\tt{B}'s behavior:
\begin{enumerate}[i)]
\item $\mathcal{M}$ takes up a position between \tt{A} and \tt{B} along the
    line joining the antennas at the two stations.
\item $\mathcal{M}$ receives the \it{sync} signal and generates a valid
    \it{response} with a delay
    \begin{equation}
      \bar{\tau}_\tt{BB} + \frac{\|\bm{x}_\mathcal{M} -
        \bm{x}_\tt{B}\|}{\|\bm{x}_\tt{A} - \bm{x}_\tt{B} \|}\left(\bar{\tau}_\tt{AB}^k + \bar{\tau}_\tt{BA}^l \right) 
    \end{equation}
    such that the RTT is
    $\bar{\tau}_\tt{AB}^k+\bar{\tau}_\tt{BB}+\bar{\tau}_\tt{BA}^l$, as
    \tt{A} expects.
  \item $\mathcal{M}$ records the unpredictable signal from \tt{A} and replays
    it to \tt{B} with an arbitrary delay $\tau_{\tt{AB}_\mathcal{M}}^k > 0$.
    This results in an error of approximately $\tau_{\tt{AB}_\mathcal{M}}^k$
    in the computed $\Delta \hat{t}_\tt{AB}^k$ at \tt{B}, as shown in
    (\ref{eq:error_adversary}).
\end{enumerate}

\subsubsection{Propagation time must be  irreducible to within $L$}
\label{sec:irreducible}
If there exists a channel that reduces the propagation time from \tt{A} to
\tt{B} or from \tt{B} to \tt{A} by more than $L$ as compared to the channel
used by \tt{A} and \tt{B}, then $\mathcal{M}$ can compromise synchronization
without detection.  The following attack assumes the propagation time from
\tt{A} to \tt{B} is reducible by more than $L$; a similar attack exploits the
situation in which the propagation time from \tt{B} to \tt{A} is reducible by
more than $L$.
\begin{enumerate}[i)]
\item $\mathcal{M}$ records the \it{sync} signal $s_\tt{A}$ going from
    \tt{A} to \tt{B}.
  \item $\mathcal{M}$ makes the recorded signal reach \tt{B} advanced by
    $|\tau_{\tt{AB}_\mathcal{M}}^k|$ compared to $s_\tt{A}$, where
    $\tau_{\tt{AB}_\mathcal{M}}^k < -L$.  An error of
    $\tau_{\tt{AB}_\mathcal{M}}^k$ is introduced in the time offset value
    computed at \tt{B} as shown in (\ref{eq:error_adversary}).
  \item $\mathcal{M}$ records the \it{response} signal $s_\tt{B}$, which has
    the expected prearranged layover time
    $\tau_\tt{BB} \approx \bar{\tau}_\tt{BB}$.  $\mathcal{M}$ replays this
    signal to \tt{A} with a delay of
    $\tau_{\tt{BA}_\mathcal{M}}^l = -\tau_{\tt{AB}_\mathcal{M}}^k$ such that
    the RTT is consistent with what \tt{A} expects.
\end{enumerate}

\subsubsection{RTT known to and measurable by \tt{A} to within $L$}
Synchronization can be compromised without detection if $|z_{\rm RTT}^{kl} -
\bar{\tau}_{\rm RTT}^{kl}| > L$ with non-negligible probability even in the
absence of an adversary.  This condition can be met if \it{a)} the prior
estimates $\bar{\tau}_\tt{AB}^k$, $\bar{\tau}_\tt{BA}^l$, or
$\bar{\tau}_\tt{BB}$ are not accurate to the corresponding true values to
within $L$, or \it{b)} the magnitude of the measurement error sum $|w_\tt{AB}^k
+ w_\tt{BA}^l|$ is larger than $L$. Note that the condition $|w_\tt{AB}^k| > L$
compromises synchronization even absent an adversary.  An adversary
$\mathcal{M}$ can exploit the condition $|z_{\rm RTT}^{kl} - \bar{\tau}_{\rm
RTT}^{kl}| > L$  as follows:
\begin{enumerate}[i)]
\item $\mathcal{M}$ initially transmits a replica of $s_\tt{A}$ such that
    \tt{B} receives identical signals from both \tt{A} and $\mathcal{M}$.
    Subsequently, $\mathcal{M}$ introduces a delay
    $\tau_{\tt{AB}_\mathcal{M}}^k > 0$ in the replayed signal $s_\mathcal{M}$.
    As assumed earlier, $s_\mathcal{M}$ exerts control and introduces an error
    of approximately $\tau_{\tt{AB}_\mathcal{M}}^k$ in the computed $\Delta
    \hat{t}_\tt{AB}^k$ at \tt{B}, as shown in (\ref{eq:error_adversary}).
\item Station \tt{B} transmits the \it{response} signal with the prearranged
    layover time $\tau_\tt{BB} \approx \bar{\tau}_\tt{BB}$ with respect to the
    delayed signal.
\item In the received signal $r_\tt{A}$, \tt{A} identifies the expected feature
    $l(k)$. The RTT, if measurable, includes the delay
    $\tau_{\tt{AB}_\mathcal{M}}^k$ introduced by $\mathcal{M}$.
  \item However, \tt{A} is unable to definitively declare an attack, since the
    errors in the modeled RTT and/or the measurement of RTT are possibly
    larger than $L$. In other words, it is not possible to claim that
    $|z_{\rm RTT}^{kl} - \bar{\tau}_{\rm RTT}^{kl}| > L$ only in
    the presence of adversarial delay.
\end{enumerate}

\section{Proof of Sufficiency}
\label{sec:sufficient}
This section presents a sufficiency proof for the set of security conditions
proposed in the previous section. A sufficiency proof guarantees secure
synchronization under the considered system and attack models. This paper draws
inspiration from the literature on modern cryptography and formalizes the
problem of secure clock synchronization with explicit definitions, assumptions,
and proofs.
\subsection{Assumptions}
\label{sec:assumptions}
This proof assumes that the system under consideration strictly complies
with the set of necessary security conditions. Specifically,
\begin{enumerate}
\item Both \tt{A} and \tt{B} use an authenticated encryption scheme to generate
    unpredictable and verifiably authentic signals in the presence of a
    probabilistic polynomial time (PPT) adversary.
\item The difference between the RTT along the communication channel between
    \tt{A} and \tt{B} and the shortest possible RTT is negligible as
    compared to $L$.
\item The difference between the modeled delays $\bar{\tau}_\tt{AB}^k$ and
    $\bar{\tau}_\tt{BA}^l$ and the true delays $\tau_\tt{AB}^k$ and
    $\tau_\tt{BA}^l$, respectively, is negligible as compared to $L$.
    \begin{equation}
        | \bar{\tau}_\tt{AB}^k - \tau_{\tt{AB}_\mathcal{N}}^k | \ll L
        \label{eq:errorAB_ll_L}
    \end{equation}
    and
    \begin{equation}
        | \bar{\tau}_\tt{BA}^l - \tau_{\tt{BA}_\mathcal{N}}^l | \ll L
        \label{eq:errorBA_ll_AL}
    \end{equation}
    Furthermore, \tt{A} and \tt{B} agree upon a fixed layover time
    $\bar{\tau}_\tt{BB}$, and the difference between this and the true layover
    time is negligible:  $|\tau_\tt{BB} - \bar{\tau}_\tt{BB}| \ll L$.
\item The standard deviation of the noise corrupting the measurements
    $t_\tt{B}^{\tt{A}_k}$ and $t_\tt{A}^{\tt{B}_l}$ is negligible compared
    to the alert limit:
    \begin{equation}
        \sigma_\epsilon \ll L
        \label{eq:noise_ll}
    \end{equation}
\end{enumerate}
Notice that the above assumptions are the same as the necessary conditions in
Section \ref{sec:necessary}, but with stricter upper bounds on the conditions.

If symmetric keys are exchanged prior to synchronization, then private-key
cryptographic schemes such as Encrypt-then-MAC \cite{bellare2000authenticated}
can be used for authenticated encryption. Alternatively, if the keys must be
exchanged over a public channel, then digital signatures
\cite{goldwasser1988digital} can be used to authenticate the encrypted
messages.  Cryptographic authentication schemes like MAC and digital signatures
generate a tag associated with a message. Qualitatively, a MAC or digital
signature scheme is secure if a PPT adversary, even when given access to
multiple valid message-tag pairs of its own choice (as many as possible in
polynomial time), cannot generate a valid tag for a new message with
non-negligible probability.  Irrespective of the cryptographic scheme used,
this proof assumes that the probability of $\mathcal{M}$ generating a new valid
\it{sync} or \it{response} signal is a negligible function of the key length
$n$:
\begin{equation}
    \mathbb{P} \left[ \mathsf{Valid} \right] < \mathsf{negl}(n)
    \label{eq:crypto}
\end{equation}
To detect an attack before the synchronization error exceeds $L$, \tt{A} must
select a threshold lower than $L$ beyond which an attack is declared. Consider
the modeled RTT, $\bar{\tau}_{\rm RTT}^{kl}$, as defined in
(\ref{eq:tRTT_bar}), and the measurement $z_{\rm RTT}^{kl}$ as
defined in (\ref{eq:RTT}). A threshold less than $L$, say $L-\delta$ with
$0 < \delta < L$, is set by station \tt{A} such that if
$|z_{\rm RTT}^{kl} - \bar{\tau}_{\rm RTT}^{kl}| > L-\delta$, then
an attack is declared.

\subsection{Definitions}
\begin{definition}
    A PPT adversary $\mathcal{M}$ succeeds if clock synchronization is
    compromised (Definition~\ref{def:compromised}) and 
    \[
        |z_{\rm RTT}^{kl} - \bar{\tau}_{\rm RTT}^{kl}| \leq L - \delta
    \]
    \label{def:success}
\end{definition}

\begin{definition}
    Faster-than-light (superluminal) propagation is defined to be hard if
    $\mathcal{M}$ cannot propagate a signal at a speed higher than the speed of
    light with non-negligible probability. Under hardness of superluminal
    propagation
    \begin{equation*}
        \mathbb{P}[\mathsf{Superluminal}] \approx 0
    \end{equation*}
\end{definition}

\begin{definition}
    A clock synchronization protocol is defined to be secure if, under the
    hardness of superluminal propagation assumption,
    \[
        \mathbb{P}[\mathsf{Success}] < \mathsf{negl}(n)
    \]
    where $\mathsf{Success}$ for $\mathcal{M}$ is defined in
    Definition~\ref{def:success}.
\end{definition}

\subsection{Proof}
In the presence of an adversary $\mathcal{M}$, the measurement $z_{\rm RTT}^{kl}$
is modeled as
\begin{align}
    z_{\rm RTT}^{kl} &= \tau_{\tt{AB}_\mathcal{N}}^k + \tau_{\tt{AB}_\mathcal{M}}^k + \tau_{\tt{BA}_\mathcal{N}}^l + \tau_{\tt{BA}_\mathcal{M}}^l + \tau_\tt{BB} + w_\tt{BA}^l \label{eq:tRTT_meas}
\end{align}
Let $\tilde{\tau}_{\tt{AB}_\mathcal{N}}^k$ and
$\tilde{\tau}_{\tt{BA}_\mathcal{N}}^l$ denote the error in the modeled
signal delay due to natural/physical phenomenon. Also, let
$\tilde{\tau}_\tt{B}$ be the difference between the intended layover time
$\bar{\tau}_\tt{BB}$ and the actual layover time $\tau_\tt{BB}$. Note that
these might be positive or negative.
\begin{align}
    \tilde{\tau}_{\tt{AB}_\mathcal{N}}^k &= \tau_{\tt{AB}_\mathcal{N}}^k - \bar{\tau}_\tt{AB}^k \label{eq:residualAB} \\
    \tilde{\tau}_{\tt{BA}_\mathcal{N}}^l &= \tau_{\tt{BA}_\mathcal{N}}^l - \bar{\tau}_\tt{BA}^l \label{eq:residualBA} \\
    \tilde{\tau}_\tt{BB} &= \tau_\tt{BB} - \bar{\tau}_\tt{BB} \label{eq:residualBB}
\end{align}
From (\ref{eq:tRTT_bar}), (\ref{eq:tRTT_meas}), (\ref{eq:residualAB}),
(\ref{eq:residualBA}), and (\ref{eq:residualBB}) it follows that
\begin{align*}
    z_{\rm RTT}^{kl} &= \bar{\tau}_{\rm RTT}^{kl} + \tilde{\tau}_{\tt{AB}_\mathcal{N}}^k + \tau_{\tt{AB}_\mathcal{M}}^k + 
                    \tilde{\tau}_{\tt{BA}_\mathcal{N}}^l + \tau_{\tt{BA}_\mathcal{M}}^l + \tilde{\tau}_\tt{BB} + w_\tt{BA}^l
\end{align*}
Following the assumptions in (\ref{eq:errorAB_ll_L}) and
(\ref{eq:errorBA_ll_AL}), the residual delays are negligible in comparison to
$L$: 
\begin{align}
    | \tilde{\tau}_{\tt{AB}_\mathcal{N}}^k | &\ll L \label{eq:residualAB_ll} \\
    | \tilde{\tau}_{\tt{BA}_\mathcal{N}}^l | &\ll L \label{eq:residualBA_ll}
\end{align}
This assumption is reasonable since otherwise the system could not confidently
meet the accuracy requirements even in the absence of an adversary.  Also, if
$\bar{\tau}_\tt{BB}$ is a short time interval and the measurement noise
$\sigma_\epsilon$ is benign, it is reasonable to assume that
\begin{align}
    | \tilde{\tau}_\tt{BB} | \ll L \label{eq:residualBB_ll}
\end{align}
Note that $\mathcal{M}$ can advance the signal by (\it{a}) forging a valid
message/tag pair, or (\it{b}) propagating the signal faster-than-light. The
assumptions of secure MAC and hardness of superluminal propagation enforce that
\begin{align*}
    \mathbb{P}[\tau_{\tt{AB}_\mathcal{M}}^k < 0] &< \mathbb{P}[\mathsf{Valid}] + \mathbb{P}[\mathsf{Superluminal}] \\
    &\approx \mathsf{negl}(n)
\end{align*}
In order to stay undetected, the adversary must ensure
\begin{align}
    L-\delta &\geq |z_{\rm RTT}^{kl} - \bar{\tau}_{\rm RTT}^{kl}| \notag \\
    &= |\tilde{\tau}_{\tt{AB}_\mathcal{N}}^k + \tau_{\tt{AB}_\mathcal{M}}^k + \tilde{\tau}_{\tt{BA}_\mathcal{N}}^l +
        \tau_{\tt{BA}_\mathcal{M}}^l + \tilde{\tau}_\tt{BB} + w_\tt{BA}^l| \label{eq:stealth}
\end{align}
At the same time, in order to compromise time transfer, from
(\ref{eq:compromised}), $\mathcal{M}$ must ensure
\begin{align}
    L &\leq |\tilde{\tau}_{\tt{AB}_\mathcal{N}}^k + \tau_{\tt{AB}_\mathcal{M}}^k + w_\tt{AB}^k| \notag \\
    &\leq |\tilde{\tau}_{\tt{AB}_\mathcal{N}}^k + w_\tt{AB}^k| + |\tau_{\tt{AB}_\mathcal{M}}^k| \notag \\
    \Rightarrow |\tau_{\tt{AB}_\mathcal{M}}^k| &\geq L - |\tilde{\tau}_{\tt{AB}_\mathcal{N}}^k + w_\tt{AB}^k| \label{eq:harm}
\end{align}
The probability of success for $\mathcal{M}$ is evaluated as
\begin{align}
  \mathbb{P}[\mathsf{Success}] &= \mathbb{P}[(\mathsf{Success}) \cap (\tau_{\tt{AB}_\mathcal{M}}^k < 0)] + \notag \\
                               &\qquad \qquad \mathbb{P}[(\mathsf{Success}) \cap (\tau_{\tt{AB}_\mathcal{M}}^k \geq 0)] \notag\\
                               &= \mathbb{P}[(\mathsf{Success}) | (\tau_{\tt{AB}_\mathcal{M}}^k < 0)] \mathbb{P}[\tau_{\tt{AB}_\mathcal{M}}^k < 0] + \notag \\
                               &\qquad \qquad \mathbb{P}[(\mathsf{Success}) \cap (\tau_{\tt{AB}_\mathcal{M}}^k \geq 0)] \notag \\
                               &\leq \mathbb{P}[\tau_{\tt{AB}_\mathcal{M}}^k < 0] \notag + \mathbb{P}[(\mathsf{Success}) \cap (\tau_{\tt{AB}_\mathcal{M}}^k \geq 0)] \notag \\
                               &< \mathsf{negl}(n) + \mathbb{P}[(\mathsf{Success}) \cap (\tau_{\tt{AB}_\mathcal{M}}^k \geq 0)] \label{eq:Asuccess}
\end{align}
In the case where $\tau_{\tt{AB}_\mathcal{M}} \geq 0$, (\ref{eq:harm})
simplifies to
\begin{equation*}
    \tau_{\tt{AB}_\mathcal{M}}^k \geq L - |\tilde{\tau}_{\tt{AB}_\mathcal{N}}^k + w_\tt{AB}^k|
\end{equation*}
Substituting the least possible value of $\tau_{\tt{AB}_\mathcal{M}}^k$
into (\ref{eq:stealth}), it follows that
\begin{equation*}
    |\tilde{\tau}_{\tt{AB}_\mathcal{N}}^k + L - |\tilde{\tau}_{\tt{AB}_\mathcal{N}}^k + w_\tt{AB}^k| + \tilde{\tau}_{\tt{BA}_\mathcal{N}}^l + 
    \tau_{\tt{BA}_\mathcal{M}}^l + \tilde{\tau}_\tt{BB} + w_\tt{BA}^l| \leq L-\delta
\end{equation*}
Notice that from the assumptions made in (\ref{eq:noise_ll}),
(\ref{eq:residualAB_ll}), (\ref{eq:residualBA_ll}), and
(\ref{eq:residualBB_ll}), all terms except $L$ and $\tau_{\tt{BA}_\mathcal{M}}^l$ on
the left-hand side of the inequality are negligible compared to $L$; thus, 
\begin{equation*}
    |L + \tau_{\tt{BA}_\mathcal{M}}^l| \leq L-\delta
\end{equation*}
Since both $L$ and $L-\delta$ are defined to be positive, the above
inequality simplifies to
\begin{equation*}
    \tau_{\tt{BA}_\mathcal{M}}^l \leq -\delta
\end{equation*}
where $\delta > 0$. Thus, for $\mathcal{M}$ to succeed in the case where
$\tau_{\tt{AB}_\mathcal{M}}^k \geq 0$, we must have that
$\tau_{\tt{BA}_\mathcal{M}}^l < 0$. As a result
\begin{equation*}
    \mathbb{P}[(\mathsf{Success}) \cap (\tau_{\tt{AB}_\mathcal{M}}^k \geq 0)] < \mathsf{negl}(n)
\end{equation*}
Thus, from (\ref{eq:Asuccess})
\begin{equation*}
    \mathbb{P}[\mathsf{Success}] < \mathsf{negl}(n)
\end{equation*}

Qualitatively, the proof presented here argues that for the adversary to
succeed, it needs to either advance the \it{sync} signal
($\tau_{\tt{AB}_\mathcal{M}} < 0$), or advance the \it{response} signal
($\tau_{\tt{BA}_\mathcal{M}} < 0$). With the use of a secure MAC (or digital
signature) and the hardness of superluminal propagation, the adversary can only
succeed with a negligible probability.

\section{Secure Constructions}
\label{sec:construction}
This section specializes the necessary and sufficient conditions for secure
clock synchronization to IEEE 1588 PTP. In addition, it presents an
alternative to PTP for wireless synchronization---a compliant synchronization
system with GNSS-like signals.
\subsection{Secure IEEE 1588 PTP}
\label{sec:secure_1588}
The necessary and sufficient conditions for secure synchronization, as adapted
to IEEE 1588 PTP, are as follows:
\begin{enumerate}
\item Stations \tt{A} and \tt{B} must use an authenticated encryption scheme to
    prevent $\mathcal{M}$ from generating valid message/tag pairs.
\item The difference between the path delays between \tt{A} and \tt{B} and the
    shortest possible path delays must be negligible as compared to $L$. For
    wireless PTP \cite{mahmood2011towards, cooklev2007ptpWLAN}, this
    implies communicating over the LOS channel as $L \rightarrow 0$. For traditional wireline PTP,
    \tt{A} and \tt{B} must attempt to communicate over the (nearly) shortest
    possible path.
\item The path delay, which is usually estimated from the RTT measurements,
    must be accurately known \it{a priori} for secure synchronization. The RTT
    measurements must be verified against the expected RTT. This
    implies that the layover time $\bar{\tau}_\tt{BB}$ must also be known to
    \tt{A}.
\end{enumerate}
Note that in the usual PTP formulation, the path delay is measured and used by
the time seeker \tt{B}. To this end, in the usual formulation \tt{A} sends the
transmit time of the \emph{sync} message and the receipt time of the
\emph{delay\_req} message (in PTP parlance). Similar conventions may be
accommodated in the system model presented in this paper, wherein \tt{A} sends
the values of $t_\tt{A}^{\tt{A}_k}$, $z_\tt{A}^l$, and $\bar{\tau}_{\rm
RTT}^{kl}$ to \tt{B}, and the following calculations may be performed and used at
\tt{B}. However, this would only be a cosmetic change and does not affect the
arguments in this paper.

\begin{figure}[ht!]
    \centering
    \includegraphics[width=8.5cm]{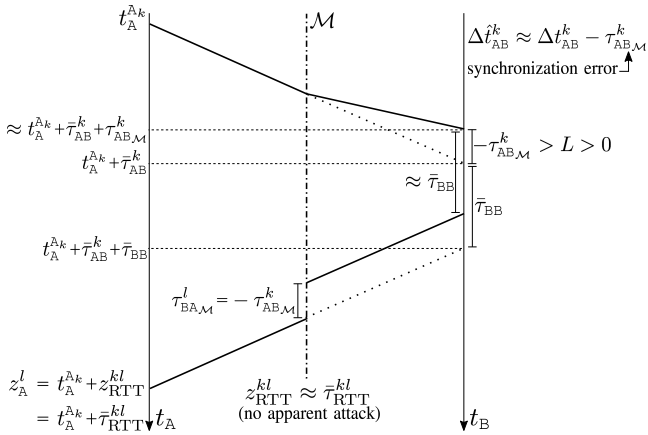}
    \caption{Illustration of an example attack against a PTP implementation that violates the second necessary condition.}
    \label{fig:shortest_path_attack}
\end{figure}

The first security condition has already been proposed in the IEEE 1588-2008
standard.  The second condition, however, has not been considered in any of the
earlier works in the literature. Following the depiction of \emph{sync} and
\emph{response} signal exchange in Fig.~\ref{fig:time_transfer} and the
attack strategy outlined in Section~\ref{sec:irreducible},
Fig.~\ref{fig:shortest_path_attack} illustrates an example attack against a
PTP implementation that does not satisfy the second necessary condition. Notice
that the existence of a shorter time path enables $\mathcal{M}$ to advance the
\emph{sync} signal relative to the authentic message from \tt{A}.
Subsequently, $\mathcal{M}$ is able to undo the effect of the advance on the
RTT by delaying the \emph{response} signal from \tt{B} to \tt{A}. Station
\tt{A} does not measure any abnormality in the RTT, and thus cannot raise an
alarm. Meanwhile, synchronization has been compromised at \tt{B}.

\begin{figure}[ht!]
    \centering
    \includegraphics[width=8.5cm]{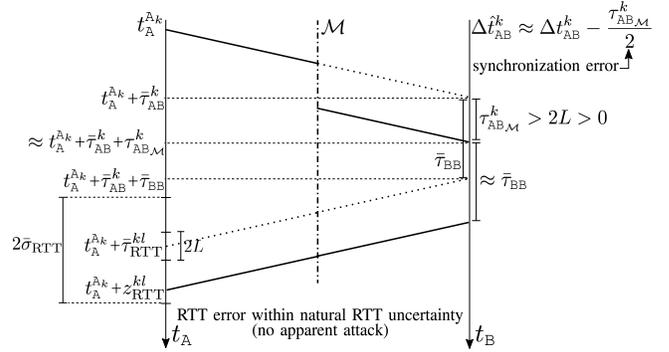}
    \caption{Illustration of an example attack against a PTP implementation that violates the third necessary condition.}
    \label{fig:rtt_attack}
\end{figure}

The third condition is similar to the proposal in \cite{ullmann2009delay} of measuring the path delays
during initialization and monitoring the delays during normal operation. However, \cite{ullmann2009delay} requires that \tt{B}
respond to \tt{A} with zero delay during initialization to enable measurement
of the reference delays. This condition is sufficient, but not necessary for
secure synchronization. The system is in fact secure even if \tt{B} is allowed
a fixed layover time.  Fig.~\ref{fig:rtt_attack} illustrates an example
attack against a PTP implementation in violation of the third necessary
condition. Note that the uncertainty of the \emph{a priori} estimate of the RTT
($\bar{\sigma}_{\rm RTT}$) is larger than the alert limit, violating the third
necessary condition which requires that the expected RTT be known to within the
alert limit (and with much higher accuracy for provable sufficiency). Even
though the measured RTT in this case is inconsistent with the expected RTT, it
cannot be definitively flagged as an attack since benign variations in the RTT
may also have led to the observed RTT.

Interestingly, at first sight, the third security condition in this paper does not resemble the
proposed defense in \cite{annessi2017securetime} that enforces an upper bound
on the synchronization error accumulated between \it{sync} messages and
recommends that \tt{B} send its timestamps to \tt{A} periodically for
verification. As explained next, this condition is in fact equivalent to the condition
of known and measurable RTT, when adapted according to the system model
considered in \cite{annessi2017securetime}.

Note that the requirement of a zero delay in \cite{ullmann2009delay}, or a
short layover time in this paper, enables \tt{A} to measure the RTT since the
transmit time of the $l$th feature in $s_\tt{B}$, that is
$t_\tt{B}^{\tt{B}_l}$, can be approximately traced back to \tt{A}'s clock to
within the alert limit as $t_\tt{A}^{\tt{A}_k} + \bar{\tau}_\tt{AB}^k +
\bar{\tau}_\tt{BB}$.  Enforcing the synchronization error to within $L$ and
transmitting \tt{B}'s timestamp to \tt{A} achieves the same objective for the
defense in \cite{annessi2017securetime}, since the transmit time from \tt{B}
can be traced back to \tt{A}'s clock with the assumed approximate
synchronization. Therefore, the proposed countermeasures in
\cite{ullmann2009delay} and \cite{annessi2017securetime} are two different
incarnations of the third security condition proposed in this paper. Of course,
the failure of both \cite{ullmann2009delay} and \cite{annessi2017securetime} to
address the second necessary condition makes their proposed defenses vulnerable
to an adversary that can communicate along a shorter time path between \tt{A}
and \tt{B}.

\subsection{Alternative Compliant System}
\label{sec:example}
This section describes an alternative wireless clock synchronization protocol that
satisfies the set of necessary and sufficient conditions presented in Section
\ref{sec:necessary}. The proposed protocol involves bi-directional exchange of
GNSS-like pseudo-random codes for continuous clock synchronization, in
contrast to discrete packet-based synchronization techniques such as NTP and
PTP.  It is offered here to illustrate the general applicability of the
proposed necessary and sufficient conditions to a range of underlying
protocols. Such a protocol can potentially be applied in two-way satellite time
transfer and terrestrial wireless clock synchronization systems for continuous
clock synchronization, in contrast to the packet-based discrete synchronization
in NTP/PTP.

The time master \tt{A} and the time seeker \tt{B} communicate wirelessly over
the LOS channel between the nodes. To simplify the analysis, it is assumed that
\tt{A} and \tt{B} securely share long sequences of pseudo-random bits prior to
synchronization. These sequences of bits will later enable generation of
unpredictable signals. The pseudo-random sequence for \tt{A} has the form
\begin{align*}
    \bm{b}_\tt{A} = \left\{ b_\tt{A}^k \right\}_{k=0}^{N}, \quad b_\tt{A}^k \in \left\{ 0,1 \right\}
\end{align*}
The pseudo-random code $C_\tt{A}(t_\tt{A})$ for \tt{A} is then generated as
\begin{align*}
    C_\tt{A}(t_\tt{A}) = 2 b_\tt{A}^k - 1 \text{ for } t_\tt{A} \in [t_\tt{A}^{\tt{A}_k}, t_\tt{A}^{\tt{A}_{k+1}}), k \in \{0,1,2,\dots\}
\end{align*}
where $t_\tt{A}^{\tt{A}_k}$ denotes the time according to \tt{A} at which the
start of the $k$th bit in \tt{A}'s signal is transmitted. The pseudo-random
nature of $\bm{b}_\tt{A}$ ensures that $C_\tt{A}(t_\tt{A})$ has good
cross-correlation properties, which enables an accurate measurement of the
time-of-arrival of \tt{A}'s signal at \tt{B}, that is, $\sigma_\epsilon \ll L$.
Station \tt{A} modulates a carrier with the code $C_\tt{A}$ and transmits a
signal $s_\tt{A}(t_\tt{A})$ whose complex baseband representation is given as
\begin{equation*}
    s_\tt{A}(t_\tt{A}) = C_\tt{A}(t_\tt{A}) \exp{\left( j \theta_\tt{A}(t_\tt{A}) \right)}
\end{equation*}
This signal is received at \tt{B} as
\begin{align*}
    r_\tt{B}(t_\tt{A},\tau_\tt{AB}) &= s_\tt{A}(t_\tt{A} - \tau_\tt{AB}) + w_\tt{AB}(t_\tt{A}) \\
    &= C_\tt{A}(t_\tt{A} - \tau_\tt{AB}) \exp{\left( j \theta_\tt{A}(t_\tt{A} - \tau_\tt{AB}) \right)} + w_\tt{AB}(t_\tt{A})
\end{align*}
where all symbols have their usual meanings as detailed in Section
\ref{sec:model}. Station \tt{B} captures a window $R_\tt{B}^k$ of $r_\tt{B}$
and correlates it with a local replica of $C_\tt{A}$. The result of the
correlation enables \tt{B} to detect the start of the $k$th bit of $C_\tt{A}$
in the window, and provides a measurement
\begin{align*}
    z_\tt{B}^k &= t_\tt{B}^{\tt{A}_k} + w_\tt{AB}^k
\end{align*}
of the time-of-arrival of the $k$th bit at \tt{B}. Furthermore, the
relationship between the start of the $k$th bit and $t_\tt{A}^{\tt{A}_k}$
enables \tt{B} to infer the latter.

If a prior estimate $\bar{\tau}_\tt{AB}^k$ of $\tau_\tt{AB}^k$ is available,
then \tt{B} estimates the clock offset $\Delta t_\tt{AB}^k$ as in
(\ref{eq:time_recovery}). 

Similar to the pseudo-random sequence and code construction for \tt{A}, \tt{B}
generates its unpredictable code $C_\tt{B}(t_\tt{B})$. \tt{A} and \tt{B} agree
on a one-to-one mapping between $C_\tt{A}$ and $C_\tt{B}$ such that \tt{B}
responds with the $l$th bit of $C_\tt{B}$ on reception of the start of the
$k$th bit of $C_\tt{A}$. Furthermore, \tt{A} and \tt{B} agree that the start of
the $l$th bit of $C_\tt{B}$ will have a code-phase offset of
$\bar{\tau}_\tt{BB}$ with respect to the start of the $k$th bit of $C_\tt{A}$.
Station \tt{B} transmits the \it{response} signal as
\begin{align*}
    s_\tt{B}(t_\tt{B}) &= C_\tt{B}(t_\tt{B}) \exp{\left( j \theta_\tt{B}(t_\tt{B}) \right)}
\end{align*}
such that
\begin{align*}
    t_\tt{B}^{\tt{B}_l} &= z_\tt{B}^k + \bar{\tau}_\tt{BB}
\end{align*}
according to the time at \tt{B}. In true time, the epoch $t_\tt{B}^{\tt{B}_l}$
corresponds to
\begin{align*}
    t_\tt{B}^{\tt{B}_l} \rightleftharpoons t_\tt{A}^{\tt{A}_k} + \tau_\tt{AB}^k + w_\tt{AB}^k + \tau_\tt{BB}
\end{align*}
Station \tt{A} receives the \it{response} as
\begin{align*}
    r_\tt{A} &= s_\tt{B}(t_\tt{B} - \tau_\tt{BA}) + w_\tt{BA}(t_\tt{A})
\end{align*}
and captures a window of the signal $R_\tt{A}^l$. \tt{A} correlates
$R_\tt{A}^l$ with a local replica of $C_\tt{B}$ to detect the start of the
$l$th bit of $C_\tt{B}$. This enables \tt{A} to measure the time-of-arrival
\begin{align*}
    z_\tt{A}^l &= t_\tt{A}^{\tt{B}_l} + w_\tt{BA}^l
\end{align*}
Moreover, the detection of the $l$th bit indicates that it was transmitted in
response to the receipt of the start of the $k$th bit of $C_\tt{A}$. Since
\tt{A} knows the start time of the $k$th bit as $t_\tt{A}^{\tt{A}_k}$, it
measures the RTT as described in (\ref{eq:RTT}).

Note that the exchange of one-time pad sequences enables the proposed system to
satisfy the first security condition. Wireless LOS communication satisfies the
second security condition, and the knowledge of the code-phase layover offset
enables \tt{A} to make an accurate prior estimate of the RTT within the alert
limit, thereby satisfying the third security condition. Thus, the proposed
system complies with all three necessary and sufficient conditions for secure
clock synchronization.

\section{System Simulation}
\label{sec:simulation}
This section presents a simulation study of a secure clock synchronization
model operating over a simplistic channel model. Unlike the abstract treatment
of delays in the security derivations presented earlier, the simulation is
carried out with models of delays experienced by the synchronization messages
over a real channel. This study also expounds the interplay between slave clock
stability, security requirements, attack models, and attack detection
thresholds that must be determined in a practical synchronization system. The
channel and attack models developed in this simulation are not comprehensive.
Rather, relatively simple models are considered to clearly demonstrate the
underlying principles. More sophisticated channel and attack models can
similarly be analyzed by following the outline of this simulation.

\subsection{Channel Model}
\begin{figure}[ht!]
    \centering
    \includegraphics[width=8.5cm]{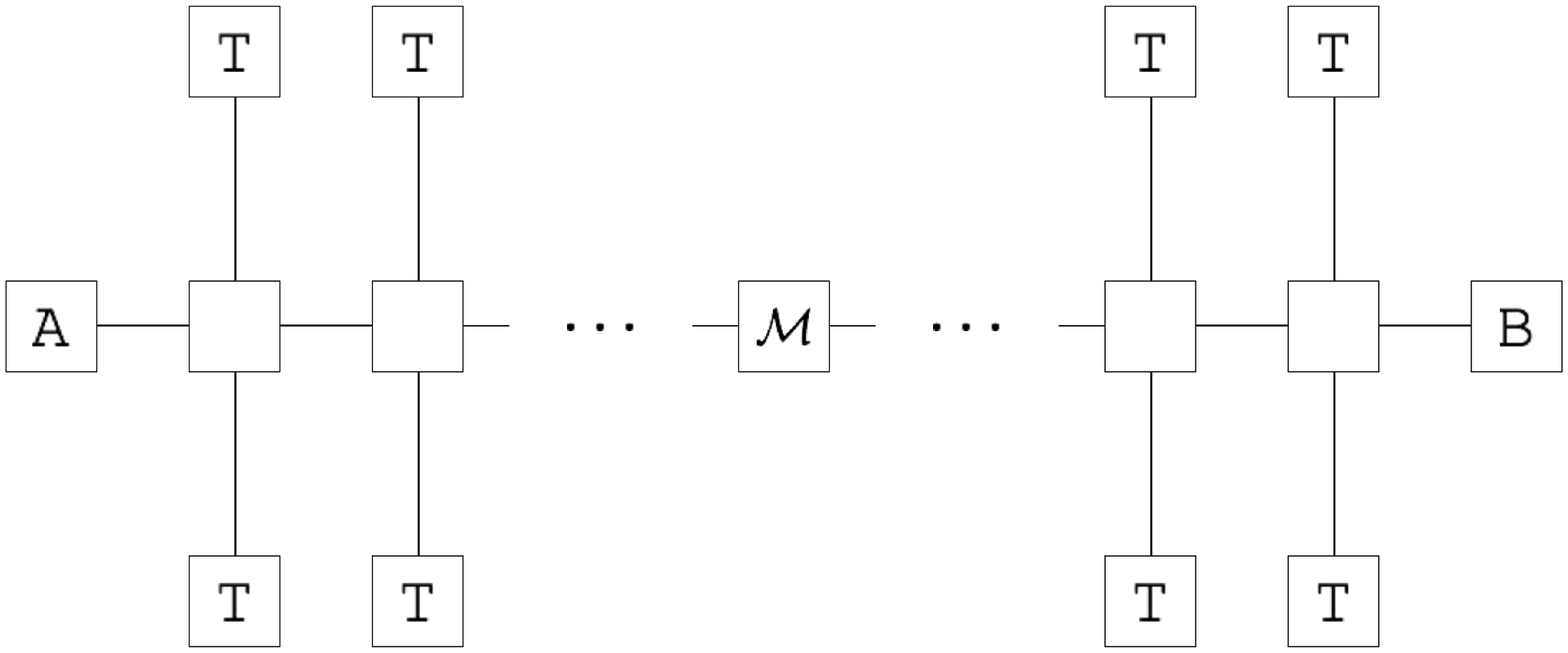}
    \caption{Schematic diagram of the network topology considered in this section.}
    \label{fig:network}
\end{figure}

The simulated system resembles a traditional local area network, and is
schematically depicted in Fig.~\ref{fig:network}. As before, \tt{A} and
\tt{B} are the time master and seeker stations, respectively. The messages
between these stations pass through a series of $N$ routers. Each router is
under network traffic loading generated by the nodes labeled \tt{T}. The routers
perform simple packet forwarding, i.e., no cryptographic operations or complex
payload modifications are performed.  Each router transmits the queued packets
at a service rate of 1 Gbps.  Each network packet is assumed to have a size of
1542 bytes. The MITM adversary $\mathcal{M}$ maliciously inserts itself along
the communication path between \tt{A} and \tt{B}.

The \emph{sync} and \emph{response} packets from \tt{A} and \tt{B} experience
processing and queueing delay at each router, and propagation/link
delay between routers. Queueing delay is the duration for which the packet is
buffered in the router before it can be transmitted. Processing delay is the
time taken by the router to process the packet header, for example, to
determine the packet's destination.  Since the routers in this simulation
perform simple packet forwarding, the processing delay is negligible as
compared to the queueing delay~\cite{ramaswamy2004characterizing}. The
propagation/link delay is also insignificant for local networks because the
propagation speed is a comparable fraction of the speed of light.  Thus, only
the queueing delay significantly contributes to the overall channel delay
variations.

Let the network idle probability for a particular router, denoted by $\rho$, be
defined as the probability of the router queue being empty at a randomly chosen
time instant. Since the synchronization packets are delay-sensitive, the
routers in this simulation implement non-preemptive priority scheduling for
synchronization packets when the queue is not empty. This means that on arrival
of a \emph{sync} or \emph{response} packet, the router is allowed to complete
the transmission of the data packet currently being serviced, if any, but is
required to service the delay-sensitive packet before the other network data in
the queue. Since the time period between consecutive
\emph{sync}-\emph{response} pairs is quite large as compared to the RTT for a
given pair, it is assumed that a router never has more than one delay-sensitive
packet in its queue. Under such scheduling, the delay experienced by the timing
messages is best modeled as follows: with probability $\rho$, the total router
delay is zero, and with probability ($1-\rho$) the total router delay is
uniformly distributed between zero and the maximum time to service a packet of
length 1542 bytes ($1542 \times 8 \times 2^{-30} \approx 11.49$ microseconds
for a Gigabit router).

\begin{figure}[ht!]
    \centering
    \includegraphics[width=8.5cm,viewport=70 210 510 390,clip=true]{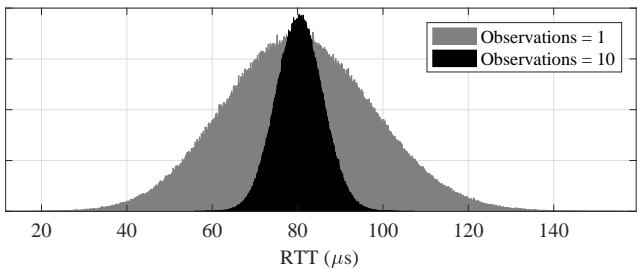}
    \caption{Empirical distribution of the RTT of \emph{sync}-\emph{response}
    pairs through a network of $N=10$ routers with network idle probability of
$\rho=0.3$. The light-shaded histogram shows the empirical distribution of the
RTT of a single \emph{sync}-\emph{response} pair. The dark-shaded histogram
shows the corresponding distribution for the mean of batches of $10$
observations of the RTT.}
    \label{fig:distribution}
\end{figure}

Given the above channel specifications and values for $N$ and $\rho$, it is
possible to perform a Monte Carlo simulation to obtain the anticipated RTT
$\bar{\tau}_{\rm RTT}$, which is taken to be the empirical mean of the RTT
measurements in the simulation, and the associated standard deviation
$\bar{\sigma}_{\rm RTT}$. As shown in Fig.~\ref{fig:distribution}, in case of a
single \emph{sync}-\emph{response} pair measurement, the RTT has an empirical
mean of $80.34$ microseconds and an empirical standard deviation of $17.09$
microseconds with $N=10$ and $\rho=0.3$. Observe that even for a relatively
small $N$, the empirical distribution approaches the Gaussian shape, but has
slightly heavier tail on the higher end of the delay.  The distribution for
mean of batches of $10$ observations has a smaller empirical standard deviation
of $5.41$ microseconds.

\subsection{System and Security Requirements}
The clock at the time seeker \tt{B} drifts with respect to the true time clock
at \tt{A} unless corrected by a \emph{sync} message from \tt{A}. As before, let
$L$ denote the alert limit for the system. Let $T$ denote a time duration over
which a perfectly synchronized clock at \tt{B} at the beginning of the
duration, absent an adversary, drifts more than $L_\mathcal{N}$ for some
$L_\mathcal{N} < L$ with a probability smaller than an acceptably small bound
$\mathbb{P}_\epsilon$.

In the system under simulation, the clock offset for \tt{B} is estimated and
corrected for every $T$ seconds. By definition of $T$, it holds that if the
clocks at \tt{A} and \tt{B} are perfectly synchronized after every $T$ seconds,
then the natural drift envelope of \tt{B}'s clock does not exceed $L$ with an
unacceptably high probability. Define
\[
    L_\mathcal{M} \triangleq L - L_\mathcal{N}
\]
Observe that if an adversary is able to introduce a synchronization error
larger than $L_\mathcal{M}$, then the system is compromised since the natural
drift of the clock at \tt{B} could potentially lead to a clock offset greater
than $L$ before the next synchronization interval, with a probability greater
than $\mathbb{P}_\epsilon$. Thus, \tt{A} must flag any adversarial delay
greater than $L - L_\mathcal{N}$ with probability higher than a desired
detection probability, denoted by $\mathbb{P}_{\rm D}$. It is worth noting that
this practical complication of the magnitude of $L_\mathcal{N}$ was abstracted
in the sufficiency proof, where the threshold was set to $L-\delta$ for
$\delta>0$.

In general, \tt{A} makes multiple measurements of the RTT between \tt{A} and
\tt{B} over the time period $T$. As shown in Fig.~\ref{fig:distribution}, the
mean of multiple observations over $T$ has a distribution with a smaller
standard deviation as compared to that of a single observation. In the
simulated system, if no attack is detected, \tt{A} updates
$\bar{\tau}_\tt{AB}^k$ every $T$ seconds based on the empirical mean of the RTT
measurements made over that period. Note that even though
$\bar{\tau}_\tt{AB}^k$ is updated based on the measurements, no updates are
applied to $\bar{\tau}_{\rm RTT}$ and $\bar{\sigma}_{\rm RTT}$, which are
predetermined by simulation or measurements under a secure calibration campaign.

The empirical mean of the measured RTT is taken as the test statistic to
detect an attack. For the attack model detailed next, it can be shown that this
test statistic becomes optimal for large values of $N$~\cite{van2004detection}.

\subsection{Attack Model}
The synchronization system considered in this simulation complies with the
necessary security conditions presented in this paper. Consequently, the
adversary $\mathcal{M}$ is unable to advance the \emph{sync} or \emph{response}
messages, and can only increase the RTT measured by \tt{A} relative to the
authentic RTT. This simulation considers a simple adversary model that
introduces a fixed delay in the measured RTT. In order to conceal its presence
while compromising synchronization with appreciable probability, $\mathcal{M}$
introduces a delay of $L_\mathcal{M} + \xi$ seconds for some small $\xi>0$.

\begin{figure}[ht!]
    \centering
    \includegraphics[width=8.5cm,viewport=70 220 510 390,clip=true]{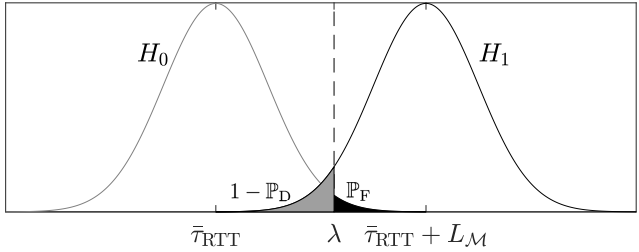}
    \caption{Representation of the distributions under $H_0$ and $H_1$ along with the detection threshold and the associated $\mathbb{P}_{\rm F}$ and $\mathbb{P}_{\rm D}$.}
    \label{fig:detection_test}
\end{figure}

Let $H_0$ denote the null hypothesis (no attack), and $H_1$ denote the
alternative hypothesis. Under $H_0$, the measured RTT at \tt{A} is drawn from
the distribution that was used to calibrate/simulate the channel delay
distribution, while under $H_1$, the measured RTT is drawn from a distribution
that is shifted from the calibration distribution by $L_\mathcal{M} + \xi$.
This is visually depicted in Fig.~\ref{fig:detection_test}. Given a detection
threshold $\lambda$, the dark-shaded region in Fig.~\ref{fig:detection_test}
denotes the probability of false alarm, $\mathbb{P}_{\rm F}$, while the
light-shaded region denotes the probability of missed detection ($1 -
\mathbb{P}_{\rm D}$). In observing Fig.~\ref{fig:detection_test}, it might be
argued, and holds true, that a reasonable attacker may introduce noise in the
introduced delay to inflate the width of the distribution under $H_1$ and
thereby decrease the probability of detection of an attack.  However, in that
case, the empirical mean test statistic is no longer optimal.
Instead, \tt{A} would incorporate the observed variance of the RTT in its test
statistic in addition to the empirical mean. In short, the attack model in this
simulation is not comprehensive, as explained previously.  For a more
sophisticated treatment of sensor deception and protection techniques, the
reader may refer to~\cite{bhatti2017hostile}.

\subsection{Simulation}
The system and attack described above have been simulated with $N=10$ and
$\rho=0.3$ for all routers. The adversarial delay $L_\mathcal{M} + \xi$ is set
to 10 microseconds, and the required probability of detection $\mathbb{P}_{\rm
D}$ is set to $0.999$. The number of RTT observations made in time $T$ are
varied between $1$ and $200$. Given the number of observations, and a required
$\mathbb{P}_{\rm D}$, the system is simulated under $H_1$ for $10^6$ detection
epochs and the maximum possible detection threshold $\lambda$ that satisfies
the detection probability is obtained. Subsequently, the system is simulated
under $H_0$ and the number of test statistics exceeding the threshold $\lambda$
are recorded. The frequency of such epochs is reported as the probability of
false alarm $\mathbb{P}_{\rm F}$.

\begin{figure}[ht!]
    \centering
    \includegraphics[width=8.5cm,viewport=70 210 510 390,clip=true]{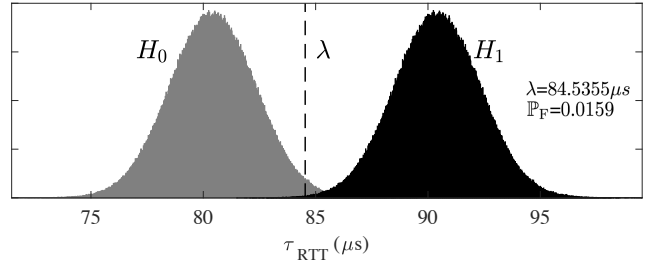}
    \caption{Distribution of the test statistic under $H_0$ and $H_1$ for $80$ RTT measurements per decision epoch. ($N=10$, $\rho=0.3$, $L_\mathcal{M}+\xi=10\mu$s, $\mathbb{P}_{\rm D}=0.999$)}
    \label{fig:detection_simulation}
\end{figure}

\begin{figure}[ht!]
    \centering
    \includegraphics[width=8.5cm,viewport=25 210 520 400,clip=true]{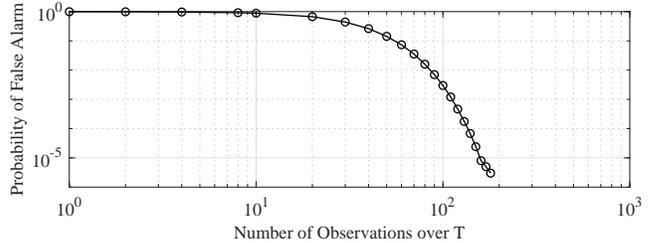}
    \caption{Probability of false alarm as a function of number of observations per decision epoch. ($N=10$, $\rho=0.3$, $L_\mathcal{M}+\xi=10\mu$s, $\mathbb{P}_{\rm D}=0.999$)}
    \label{fig:T_PF}
\end{figure}

Fig.~\ref{fig:detection_simulation} shows the above procedure for $80$ RTT
measurements made per test statistic. In this case, $\lambda$ is obtained to be
84.53 microseconds and the corresponding $\mathbb{P}_{\rm F}$ is $1.59\%$.
Fig.~\ref{fig:T_PF} shows a log-log plot of $\mathbb{P}_{\rm F}$ as a
function of the number of observations made per test statistic. When the number of
observations is greater than $160$, no false alarms were observed with $10^6$
trials. For the given channel delay variation statistics, the probability of
false alarm is very high for small number of observations per decision epoch
since the threshold $\lambda$ that must be set to detect an attack with the
required $\mathbb{P}_{\rm D}$ is large in comparison to the minimal delay that the
adversary must introduce to compromise synchronization ($L_\mathcal{M}$). For a
more stable channel, such as a wireless or PTP-aware channel, fewer
measurements per decision epoch would suffice.

\subsection{Practical Implications}
Section~\ref{sec:assumptions} makes fairly remarkable assumptions about the
synchronization system to show provably secure time transfer. For instance, it
requires that errors in the \emph{a priori} estimate of the RTT of the
timing messages be negligible compared to the alert limit.  Nonetheless, as
shown in this section, for a given channel with bounded delay variations and a
given slave clock, some level of security guarantee can be made for a
synchronization system that satisfies the necessary and sufficient conditions
presented herein.  For concreteness, consider a system that requires an alert
limit of $L=100$ microseconds and a slave clock that drifts no more than
$L_\mathcal{N}=50$ microseconds over a period of $T=1$ second with acceptably
high probability ($1 - \mathbb{P}_\epsilon$). Then, for $N=10$ and $\rho=0.3$,
if \tt{A} makes 10 RTT measurements over 1 second, the empirical mean test
statistic is distributed as the dark-shaded distribution in
Fig.~\ref{fig:distribution} with a standard deviation of $\approx 5.4$
microseconds. For $L_\mathcal{M} = L - L_\mathcal{N} = 50$ microseconds, a
threshold of $\approx \bar{\tau}_{\rm RTT} + 30$ microseconds will yield a
missed detection rate of approximately $1$ in $15000$, and a false alarm rate
of approximately $3.5$ in $1$ million. With a more stable slave clock or more
measurements per second, these probabilities can be made more favorable.

Another important concern that has not been addressed in the simulation is that
of the incorporation of cryptographic constructs in the synchronization
protocol. The encryption and decryption algorithms are often complex and take
non-negligible processing time to execute. However, note that at \tt{A}, the
\emph{sync} message is timestamped \emph{after} the encryption process, and
thus the time taken for encryption is inconsequential. At \tt{B}, it is
important to concede that the decryption of the \emph{sync} message and the
encryption of the \emph{response} message cannot be assumed to happen
instantaneously. This has been accounted for by allowing the layover time
$\bar{\tau}_\tt{BB}$ for the cryptographic processes to execute. Once again,
the receipt timestamp of the \emph{response} message at \tt{A} is applied
\emph{before} the decryption process, and hence the decryption time at \tt{A}
is inconsequential. Thus, compliance with the first security condition must not
pose significant practical challenges.

\section{Conclusions}
\label{sec:conclusion}
A fundamental theory of secure clock synchronization was developed for a
generic system model. The problem of secure clock synchronization was
formalized with explicit assumptions, models, and definitions. It was shown
that all possible one-way clock synchronization protocols are vulnerable to
replay attacks.  A set of necessary conditions for secure two-way clock
synchronization was proposed and proved. Compliance with these necessary
conditions with strict upper bounds was shown to be sufficient for secure
clock synchronization, which is a significant result for provable security in
critical infrastructure. The general applicability of the set of security
conditions was demonstrated by specializing these conditions to design a
secure PTP protocol and an alternative secure two-way clock synchronization
protocol with GNSS-like signals. Results from a simulation with models of
channel delays were presented to expound the interplay between slave clock
stability, security requirements, attack models, and attack detection
thresholds.

\section*{Acknowledgments}
This project has been supported by the National Science Foundation under Grant
No. 1454474 (CAREER), by the Data-supported Transportation Operations and
Planning Center (DSTOP), a Tier 1 USDOT University Transportation Center, and
by the U.S. Department of Energy under the TASQC program led by Oak Ridge
National Laboratory.

\bibliographystyle{ieeetran} 
\bibliography{pangea}

\begin{IEEEbiography}[{\includegraphics[width=1in,height=1.3in,clip,keepaspectratio]{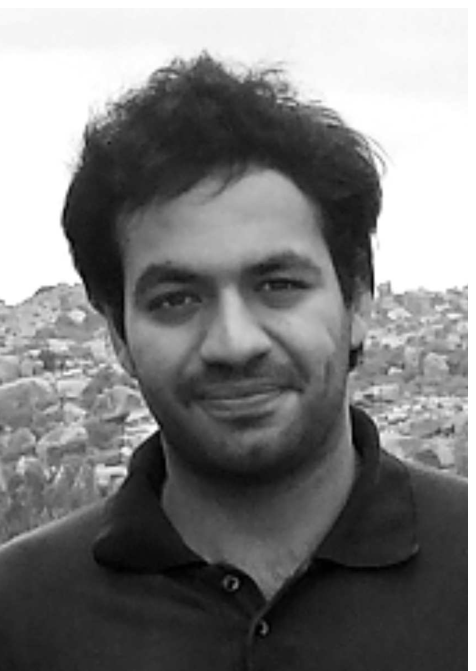}}]{Lakshay Narula}
received the B.Tech. degree in electronics engineering from IIT-BHU, India, in
2014, and the M.S. degree in electrical and computer engineering from The
University of Texas at Austin, Austin, TX, USA, in 2016.

He is currently a Ph.D. student with the Department of Electrical and Computer
Engineering at The University of Texas at Austin, and a Graduate Research
Assistant at the UT Radionavigation Lab. His research interests include GNSS signal
processing, secure perception in autonomous systems, and detection and
estimation.

Lakshay has previously been a visiting student at the PLAN Group at University
of Calgary, Calgary, AB, Canada, and a systems engineer at Accord Software \&
Systems, Bangalore, India. He was a recipient of the 2017 Qualcomm Innovation
Fellowship.
\end{IEEEbiography}
\begin{IEEEbiography}[{\includegraphics[width=1in,height=1.3in,clip,keepaspectratio]{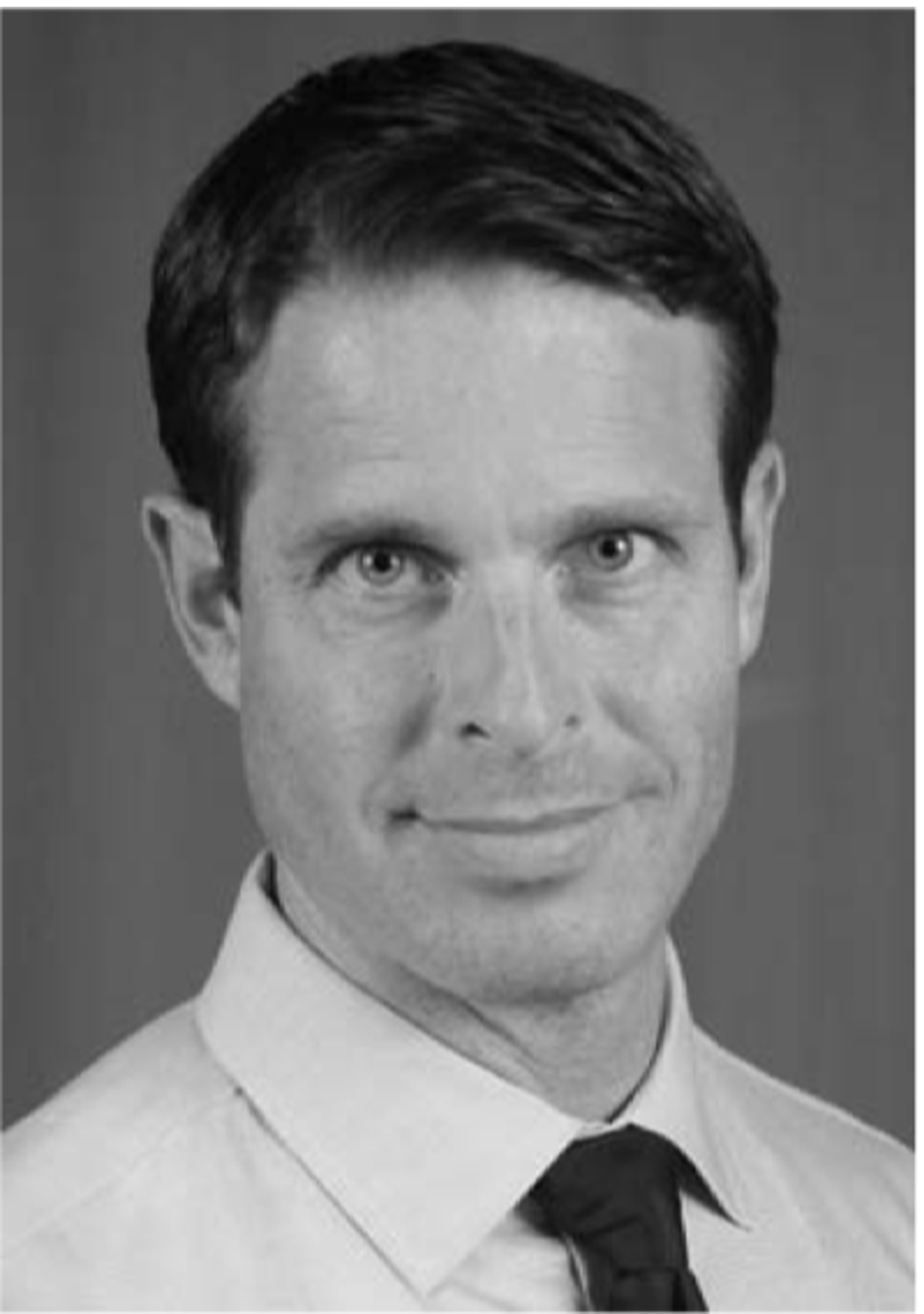}}]{Todd E. Humphreys}
received the B.S. and M.S. degrees in electrical and computer engineering from
Utah State University, Logan, UT, USA, in 2000 and 2003, respectively, and the
Ph.D. degree in aerospace engineering from Cornell University, Ithaca, NY, USA,
in 2008.

He is an Associate Professor with the Department of Aerospace Engineering and
Engineering Mechanics, The University of Texas (UT) at Austin, Austin, TX, USA,
and Director of the UT Radionavigation Laboratory. He specializes in the
application of optimal detection and estimation techniques to problems in
satellite navigation, autonomous systems, and signal processing. His recent
focus has been on secure perception for autonomous systems, including
navigation, timing, and collision avoidance, and on centimeter-accurate
location for the mass market.

Dr. Humphreys received the University of Texas Regents' Outstanding Teaching
Award in 2012, the National Science Foundation CAREER Award in 2015, and the
Institute of Navigation Thurlow Award in 2015.
\end{IEEEbiography}
\vfill

\end{document}